\documentclass[11pt]{article}
\usepackage{jheppub}

\usepackage{graphicx}
\usepackage{dcolumn}
\usepackage{bm}
\usepackage{slashed}
\usepackage{physics}
\usepackage{extarrows}
\usepackage{multirow,array}
\usepackage{amsmath,amssymb,physics}
\usepackage{subfigure}
\usepackage{enumitem}
\setlist{nolistsep}
\usepackage{xcolor}
\usepackage{url}
\usepackage{slashed}
\usepackage{hyperref}
\definecolor{nicered}{rgb}{0.5,0.,0.}
\definecolor{nicegreen}{rgb}{0.,0.5,0.}
\definecolor{niceblue}{rgb}{0.,0.,0.5}
\hypersetup{colorlinks,citecolor=nicegreen,linkcolor=nicered,urlcolor=niceblue}
\usepackage{braket}
\usepackage{colortbl}
\usepackage{gensymb}
\usepackage{comment}

\hypersetup{
  pdfauthor={Tao Han, Yang Ma, Keping Xie}
  pdftitle={Quark and Gluon Contents of a Lepton at High Energies}
  pdfkeywords={QCD, Electroweak, Parton distribution functions, colliders}
}
\newcommand{\gev}{{\ensuremath\rm GeV}}
\newcommand{\tev}{{\ensuremath\rm TeV}}

\newcommand{\bea}{\begin{equation}\begin{aligned}}
\newcommand{\eea}{\end{aligned}\end{equation}}
\newcommand{\beq}{\begin{equation}}
\newcommand{\eeq}{\end{equation}}
\newcommand{\bear}{\begin{eqnarray}}
\newcommand{\eear}{\end{eqnarray}}

\newcommand{\ee}{e^+e^-}
\newcommand{\mm}{\mu^+\mu^-}
\newcommand{\ww}{W^+ W^-}

\newcommand{\muQCD}{\mu_\textrm{QCD}}
\newcommand{\LQCD}{\Lambda_\textrm{QCD}}
\newcommand{\muEW}{\mu_\textrm{EW}}

\newcommand{\lsea}{\ell_{\textrm{sea}}}
\newcommand{\lval}{\ell_{\textrm{val}}}

\newcommand{\GeV}{~\textrm{GeV}}
\newcommand{\MeV}{~\textrm{MeV}}

\title{Quark and Gluon Contents of a Lepton at High Energies} 
\author[a]{Tao Han,}
\affiliation[a]{Pittsburgh Particle Physics, Astrophysics, and
	Cosmology Center, Department of Physics and Astronomy, University of
	Pittsburgh, Pittsburgh, PA 15260, USA
}
\author[a]{Yang Ma,}
\author[a]{and Keping Xie}

\emailAdd{than@pitt.edu}
\emailAdd{mayangluon@pitt.edu}
\emailAdd{xiekeping@pitt.edu}
\preprint{PITT-PACC-2104} 

\abstract{In high-energy leptonic collisions, such as at a multi-TeV muon collider, the collinear splittings of the electroweak (EW) gauge bosons and leptons are the dominant phenomena, and the scattering processes should thus be formulated in terms of the EW parton distribution functions (EW PDFs). We complete this formalism in the Standard Model to include the QCD sector and evaluate the quark and gluon PDFs inside a lepton at the double-log accuracy. The splittings of the photon and subsequently the quarks and gluons control the quark/gluon PDFs below the EW scale. The massive gauge bosons lead to substantial contributions at high scales. The jet production cross section can reach the order of a few nb (50 pb) in $\ee$ ($\mm$) collisions, at the TeV c.m.~energies with a moderate acceptance cut, that governs the overall event shape up to about $p_T^j \sim 60$ GeV. To complete the picture, we also provide an estimation of the total cross section for the photon-induced hadronic production at low partonic energies, which can reach the level of one hundred or a few tens of 
nb in high-energy electron or muon collisions. 
}

\begin{document}
\maketitle


\section{Introduction}
\label{sec:intro}

There have been renewed interests recently in exploring physics opportunities at multi-TeV lepton colliders, thanks to the breakthrough in the cooling technology for a muon beam \cite{Delahaye:2019omf},
and the advancement of the wake-field electron acceleration technology \cite{ALEGRO:2019alc}.
This exciting possibility could lead us to an unexplored regime at the energy and luminosity frontier for new physics reach beyond the Standard Model (SM). Indeed, beyond the extensive studies for a multi-TeV $e^+e^-$ collider of the CERN Compact Linear Collider (CLIC) \cite{Roloff:2018dqu},
some recent works on a high-energy muon collider have shown great physics potential for precision SM Higgs physics \cite{Han:2020pif,Buttazzo:2020uzc,Han:2021lnp}, BSM heavy Higgs boson discovery \cite{Bandyopadhyay:2020otm,Han:2021udl}, 
WIMP dark matter searches \cite{Han:2020uak,Capdevilla:2021fmj}, electroweak phase transition \cite{Liu:2021jyc},
lepton-universality violation \cite{Huang:2021nkl,Huang:2021biu}, and a broad coverage for other new physics scenarios \cite{Costantini:2020stv,Capdevilla:2020qel,Gu:2020ldn,Yin:2020afe,Capdevilla:2021rwo,Buttazzo:2020eyl}.

While a lepton collider has the great merit for a monochromatic energy spectrum at the designed center-of-momentum (c.m.) energy $\sqrt s$, it simultaneously offers a broad energy spectrum due to the enhanced collinear radiation of the electroweak (EW) gauge bosons. This leads to the familiar phenomena of the photon-photon collisions \cite{Budnev:1974de,Brodsky:2005wk}. In fact, the vector-boson fusion (VBF) mechanism dominates the physical processes in high-energy leptonic collisions \cite{Barger:1996kp,Costantini:2020stv,Han:2020uid}. To properly describe those reactions, it was emphasized recently \cite{Han:2020uid} that it is appropriate to adopt the partonic picture by introducing the electroweak parton distribution functions (EW PDFs) \cite{Ciafaloni:2005fm,Chen:2016wkt,Manohar:2018kfx,Fornal:2018znf}, which run according to the evolution equations of the unbroken gauge theory of SU(2)$_{\rm L}\otimes$U(1)$_{\rm Y}$ at high energies above the EW scale. It is important to formulate the EW PDFs to predict the SM expectations at the ultra-high energies, before estimating the sensitivity for new physics searches. 

In the subsequent splitting of the EW gauge bosons, quarks enter the picture of the EW partons from $\gamma/Z,W^\pm\to q\bar q'$. The strong QCD interactions of quarks and gluons take over and the coupled DGLAP equations of the full Standard Model must be invoked \cite{Bauer:2017isx,Bauer:2018arx}. This would yield QCD contributions in leptonic collisions and thus lead to new mechanisms for the production of colored states \cite{Han:2010rf}. In fact, quark contributions to QCD jet production in $\ee$ collisions were considered in the literature \cite{Drees:1994eu}. They are the dominant phenomena in the kinematical region with forward-backward scattering and lower energy transfer. It is thus important to have a clear understanding of the events and the characteristics taking into account the EW and QCD interactions of the partons in high-energy lepton collisions. Motivated by the recent discussions on the future high-energy $e^+e^-$ or $\mu^+\mu^-$ colliders, we consider a collider with the c.m.~energies  
\beq
\sqrt{s} = 3\ {\rm TeV} - 15\ {\rm TeV}, 
\eeq
with a few benchmark points as 3 TeV, 6 TeV, 10 TeV, and 14 TeV. 
The 3-TeV c.m.~energy is the benchmark for the Compact Linear Collider \cite{Roloff:2018dqu} and the higher energies are those under discussion for future muon colliders \cite{Delahaye:2019omf}. 
The total integrated luminosity is assumed to be in the range of $(1-10)$ ab$^{-1}$. 

In Sec.~\ref{sec:PDF}, we present the full DGLAP equations for the quarks and gluons coupled to the EW sector in the SM. In dealing with the full SM spectrum, the physics is characterized by two scales, namely, $\LQCD \sim 200$ MeV and $\Lambda_{\rm EW}\sim 250$ GeV. To assure perturbativity, we take $\muQCD=0.5$ GeV, inspired by the critical scale adopted in Ref. \cite{Drees:1994eu}. The different choice of $\muQCD$ is ascribed to the non-perturbative uncertainty. The EW threshold is taken at $\muEW=M_Z$ to excite the EW gauge bosons and the top quark. 
We solve DGLAP equations numerically and calculate the quark and gluon PDFs of a lepton at representative factorization scales. We find substantial quark and gluon luminosities resulting from an initial electron and a muon, especially in the relatively low invariant mass region.

After setting up the QCD/EW partonic formalism, we calculate the SM prediction for some leading production processes at high-energy electron and muon colliders as shown in Sec.~\ref{sec:SMProcesses}. 
In particular, we present in detail the QCD jet production initiated by quarks and gluons, which present the dominant contributions, up to the transverse momenta about 60 GeV.
We also provide an estimation of the total cross section for the photon-induced hadronic production at low partonic energies, which dominates the event shape in this energy regime. 
We summarize our results and conclude Sec.~\ref{sec:sum}. 

\section{The parton distribution functions for quarks and gluons}
\label{sec:PDF}

Different from a proton beam, the parton contents inside of a lepton can be calculated perturbatively. The evolutions of parton distribution functions (PDFs) over a factorization scale $Q$ are governed by the well-known DGLAP equations \cite{Altarelli:1977zs,Gribov:1972ri,Lipatov:1974qm,Dokshitzer:1977sg} 
\begin{equation}
\label{eq:DGLAP}
\frac{\dd f_i}{\dd\log Q^2}=\sum_{I}\frac{\alpha_I}{2\pi}\sum_{j}P_{i,j}^{I}\otimes f_j,
\end{equation}
where the index $I$ loops the different SM interactions. The symbol $\otimes$ stands for a convolution
\begin{equation}\label{eq:conv}
\left[f\otimes g\right](x)
=\int_0^1\dd \xi\dd\zeta\delta(x-\xi\zeta)f(\xi)g(\zeta)
=\int_x^1\frac{\dd\xi}{\xi}f(\xi)g\left(\frac{x}{\xi}\right).
\end{equation}
$P_{i,j}^I$ are the splitting functions for $j\to i$ under the SM interaction $I$, and $x$ is the momentum fraction carried by the daughter particle $i$. The leading order QCD and QED splitting functions are known for decades and can be found in textbooks \cite{Ellis:1991qj,Campbell:2017hsr}. They are extended to include mixed term $\mathcal{O}(\alpha\alpha_s)$ in Ref. \cite{deFlorian:2015ujt} and next-to-leading order (NLO) QED in Ref. \cite{deFlorian:2016gvk}. The pure QCD splittings are known up to next-to-next-to-leading order (NNLO) \cite{Vogt:2004mw,Moch:2004pa}, which are employed to determine the QCD PDFs of proton in several global fitting groups \cite{Abramowicz:2015mha,Alekhin:2016uxn,Bailey:2020ooq,Hou:2019efy,Ball:2017nwa,Jimenez-Delgado:2014twa}. The QED and QED mixed evolutions are adopted to determine the photon content in Refs.~\cite{Martin:2004dh,Ball:2013hta,Schmidt:2015zda}. Recently, a more precise determination of the photon PDF of a proton in terms of the electromagnetic structure functions was proposed as the LUXqed formulation \cite{Manohar:2016nzj,Manohar:2017eqh}, which are employed in the global PDF analysis \cite{Bertone:2017bme,Harland-Lang:2019pla,Xie:2021equ}. The splitting functions are extended to the EW theory to involve the EW gauge bosons and chiral states in Refs.~\cite{Ciafaloni:2005fm,Chen:2016wkt}, which are adopted to determine the proton EW PDFs~\cite{Bauer:2017isx,Bauer:2018arx}.

As discussed in Sec.~\ref{sec:intro}, for a leptonic beam, the DGLAP evolution equations in Eq.~(\ref{eq:DGLAP}) run differently in three regions of the physical scales.
The initial condition starts from the lepton mass, and the QED PDFs (including the photon, charged leptons, and quarks) run in terms of the QED gauge group.
Starting at $\muQCD$, the QCD interaction begins to enter. The QCD and QED evolutions run simultaneously until $\muEW$, where the complete SM sector
begins to evolve according to the unbroken SM gauge group. In such a way, we need two matchings, at $\muQCD$ and $\muEW$, respectively.\footnote{In a realistic situation, one should perform a matching whenever crossing a heavy-flavor threshold, such as at $m_\tau,m_c,m_b,m_t$. In practice, the multiple scales make the DGLAP evolution complicated, which is beyond the scope of this work. We defer the related aspects to a future dedicated study~\cite{HMX}. 
As long as the observables under consideration are not heavy-flavor sensitive and the physical scale is well above their mass thresholds, 
the heavy flavors just behave similarly to the light sea flavors that are all generated dynamically. Therefore, we treat them on the equal footing classified by the matching scales $\muQCD$ and $\muEW$, just for simplicity.} 
As the QED and QCD gauge groups conserve the charge and parity symmetry, the PDFs below $\muEW$ can be treated with no polarization, as long as the initial lepton beams are unpolarized. As pointed out already in Refs.~\cite{Bauer:2018arx,Han:2020uid}, the polarization plays an important role in the EW PDFs above the EW scale, even for the unpolarized initial beams. Consequently, the photon and gluon become polarized due to the fermion chiral interactions. 

\subsection{PDF evolution in QED and QCD}
\label{sec:QCD}
For the sake of illustration, we take the electron beam as an example. The presentation is similarly applicable to the muon beam by recognizing a different mass. 
In solving the QED and QCD DGLAP equations, it is customary to define the fermion PDFs in a basis of gauge singlets and non-singlets. The singlet PDFs can be defined as 
\beq
f_L=\sum_{i=e,\mu,\tau}(f_{\ell_i}+f_{\bar{\ell}_i}), ~ 
f_U=\sum_{i=u,c}(f_{u_i}+f_{\bar{u}_i}), ~ 
f_D=\sum_{i=d,s,b}(f_{d_i}+f_{\bar{d}_i}),
\eeq
where the subscripts refer to the fermion flavors and we have excluded the top quark below the EW scale. 
The DGLAP equations in Eq.~(\ref{eq:DGLAP}),
involving the photon and gluon, can be written as
\begin{equation}\label{eq:SG}
\frac{\dd}{\dd\log Q^2}
\begin{pmatrix}
f_L\\
f_U\\
f_D\\
f_\gamma\\
f_g
\end{pmatrix}
=
\begin{pmatrix}
P_{\ell\ell} & 0 & 0 & 2N_\ell P_{\ell\gamma} & 0\\
0 & P_{uu} & 0 & 2N_uP_{u\gamma} & 2N_uP_{ug}\\
0 & 0 & P_{dd} & 2N_dP_{d\gamma} & 2N_dP_{dg}\\
P_{\gamma\ell} & P_{\gamma u} & P_{\gamma d} & P_{\gamma\gamma} & 0\\
0 & P_{g u} & P_{gd} & 0 & P_{gg}
\end{pmatrix}
\otimes 
\begin{pmatrix}
f_L\\
f_U\\
f_D\\
f_\gamma\\
f_g
\end{pmatrix},
\end{equation}
where the active flavors below the EW scale are 
\beq
N_\ell=3, ~ N_u=2, ~N_d=3.
\label{eq:nf}
\eeq
We remind that the splitting functions $P_{q\gamma}$ and $P_{qg}$ ($q=u,d$) contain color factor implicitly.
In this work, we only consider the leading order splittings. The $P_{ij}$ defined here include the gauge couplings $\alpha$ and $\alpha_s$ in Eq.~(\ref{eq:DGLAP}), which evolve with scale as well. The initial condition for an electron beam at the leading order is 
\begin{equation}
\label{eq:init}
f_{e/e}(x,m_e^2)=f_{L}(x,m_e^2)=\delta(1-x),
\end{equation}
while all the other PDFs are zero at the initial scale $Q^2=m_e^2$.

The non-singlet PDFs can be defined as 
\bear
f_{\ell_i}^{\rm NS}&=&f_{\ell_i}-f_{\bar{\ell}_i}, ~f_{\ell,12}=f_{\bar{e}}-f_{\bar{\mu}}, ~f_{\ell,13}=f_{\bar{e}}-f_{\bar{\tau}}, \\
f_{u_i}^{\rm NS}&=&f_{u_i}-f_{\bar{u}_i}, f_{u,12}=f_u-f_c , \\
f_{d_i}^{\rm NS}&=&f_{d_i}-f_{\bar{d}_i}, ~ f_{d,12}=f_{d}-f_{s}, ~ f_{d,13}=f_{d}-f_{b}.
\eear
The DGLAP equations for the non-singlet PDFs are written as
\begin{equation}\label{eq:DGLAPNS}
\dv{}{\log Q^2}f^{\rm NS}=P_{ff}\otimes f^{\rm NS}.
\end{equation}
where $f=\ell,u,d$. 
At the starting scale $Q^2=m_e^2$, the only non-trivial non-singlet PDF is
\begin{equation}\label{eq:NSval}
f_{e}^{\rm NS}=f_{e}-f_{\bar{e}} = \delta(1-x),
\end{equation}
while all the other non-singlet PDFs are trivially zero and remain to be zero at high scales due to the zero initial conditions.

We can now construct the PDFs for each flavor in terms of the singlet and non-singlet PDFs. The valence flavor PDF is 
\begin{equation}
f_{e}=\frac{f_{L}+(2N_{\ell}-1)f_{e}^{\rm NS}}{2N_{\ell}},
\end{equation}
and the sea fermion PDFs are 
\bear
\label{eq:Lsea}
&&f_{\bar{e}}=f_{\mu}=f_{\bar{\mu}}=f_{\tau}=f_{\bar{\tau}}=\frac{f_L-f_{e}^{\rm NS}}{2N_{\ell}},\\
\label{eq:Usea}
&&f_{u}=f_{\bar{u}}=f_{c}=f_{\bar{c}}=\frac{f_{U}}{2N_u}, \\
\label{eq:Dsea}
&&f_{d}=f_{\bar{d}}=f_{s}=f_{\bar{s}}=f_{b}=f_{\bar{b}}=\frac{f_{D}}{2N_d}.
\eear

A few remarks are in order. \\
$\bullet$
We would like to remind the reader that the relations of the sea flavor PDFs in Eqs.~(\ref{eq:Lsea}-\ref{eq:Dsea}) are valid only when we ignore the fermion masses in accordance with the rigorous collinear factorization. The PDFs for heavy flavors will receive threshold corrections when their masses are taken into account, as already commented on with multiple scales. 
This would lead to finite corrections of the order $(\alpha/2\pi) \log(m_f^2/m_\ell^2)$ to the heavy-flavor PDFs~\cite{HMX}. More detailed studies for the threshold matching are beyond the scope of our current interests. \\
$\bullet$
Below $\muQCD$, the QCD confinement sets in.
As such, the picture of ``vector-meson-dominance'', {\it e.g.} $\gamma-\rho$ mixing, gives the leading contribution to the photonic interactions, as already included in most of the photon-PDFs. 
It is 
expected to be bounded by $\alpha^2\log^2(\muQCD^2/m_\ell^2)$.
In our practical treatment, we only run the QED gauge group in the DGLAP evolution. The $\gamma\to q\bar{q}$ splitting serves as a source of the initial conditions of the QCD PDFs at the matching scale $\muQCD$, similar to the quark-parton model Ans\"atze adopted in Ref.~\cite{Drees:1994eu}.
\\
$\bullet$ 
Above $\muEW$,
the unbroken SM gauge interactions come into play and the PDFs receive EW corrections. 
The EW gauge boson $W/Z$ and top-quark parton become active,\footnote{Here, we ignore the threshold correction, $\log(M_t^2/\muEW^2)$, to the top-quark PDFs, which is valid as long as the physical energy scale is far above the EW scale, \emph{i.e.}, $Q^2\gg\muEW^2$.} and the complete EW PDFs become polarized due to the chiral couplings, as outlined in a previous publication \cite{Han:2020uid}.
We will properly include the EW effects in the rest of our calculations.
%
\subsection{PDFs and partonic luminosities at a lepton collider}

%
With the formalism in the last section, we can compute the parton distribution functions of quarks and the gluon in a high-energy lepton, along with leptons and the photon. Because of the complexity of the coupled integrodifferential equations, one encounters highly technically challenging calculations, with example of non-singlet PDF of the valence lepton demonstrated in App. \ref{app:solution}.
The comprehensive details are left for a future work \cite{ours}.

\begin{figure}[tb]
\includegraphics[width=.48\textwidth]{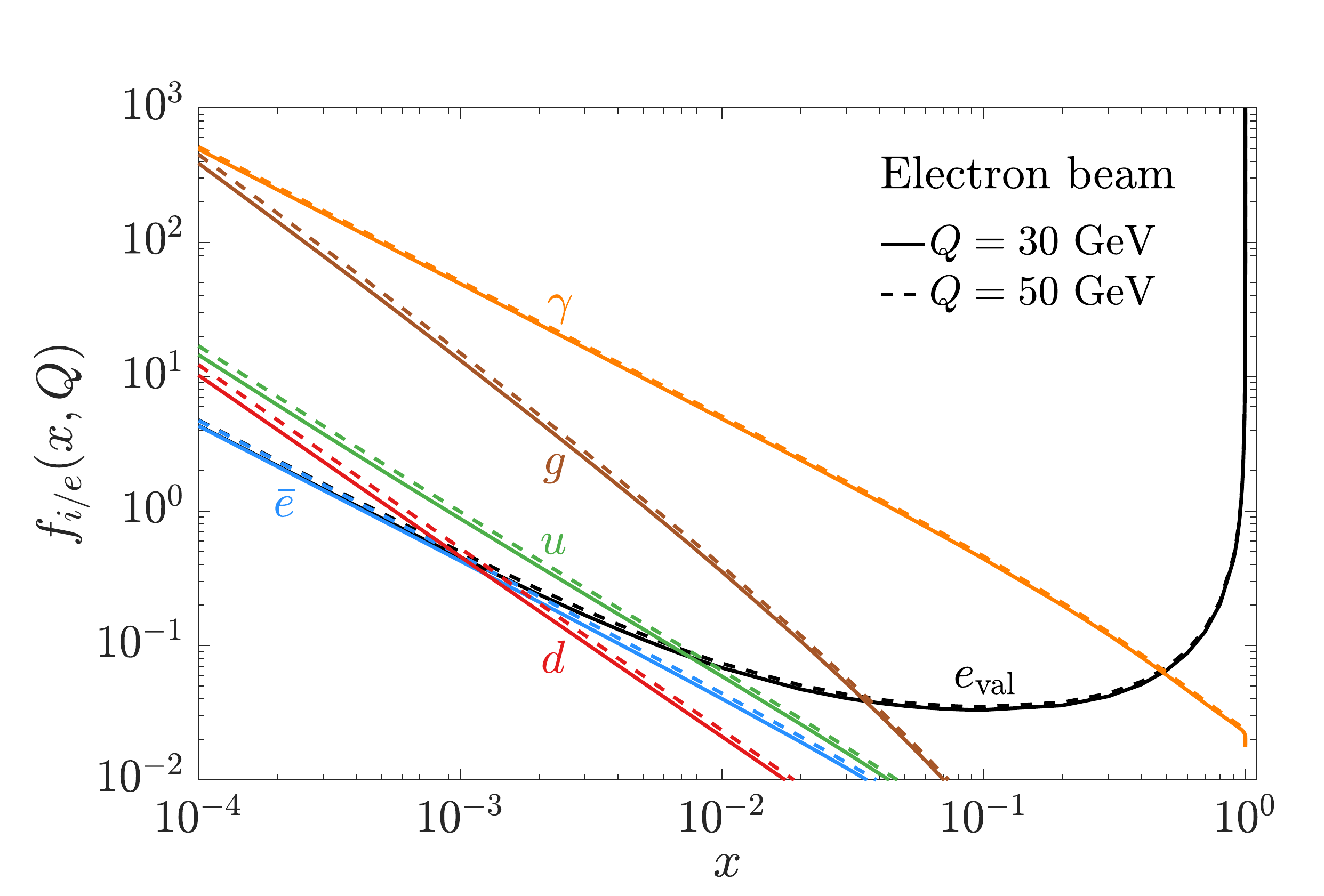}
\includegraphics[width=.48\textwidth]{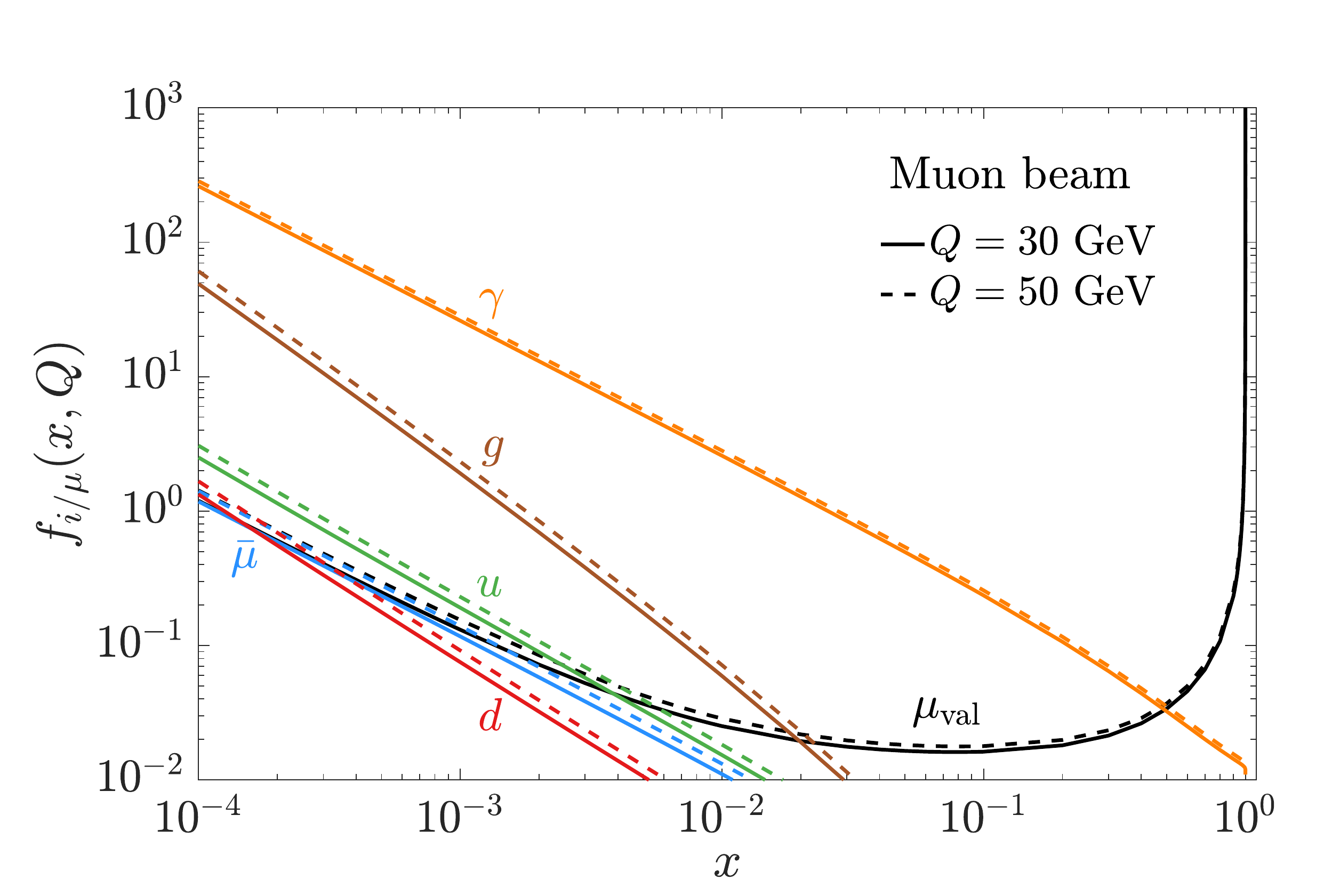}
\includegraphics[width=.48\textwidth]{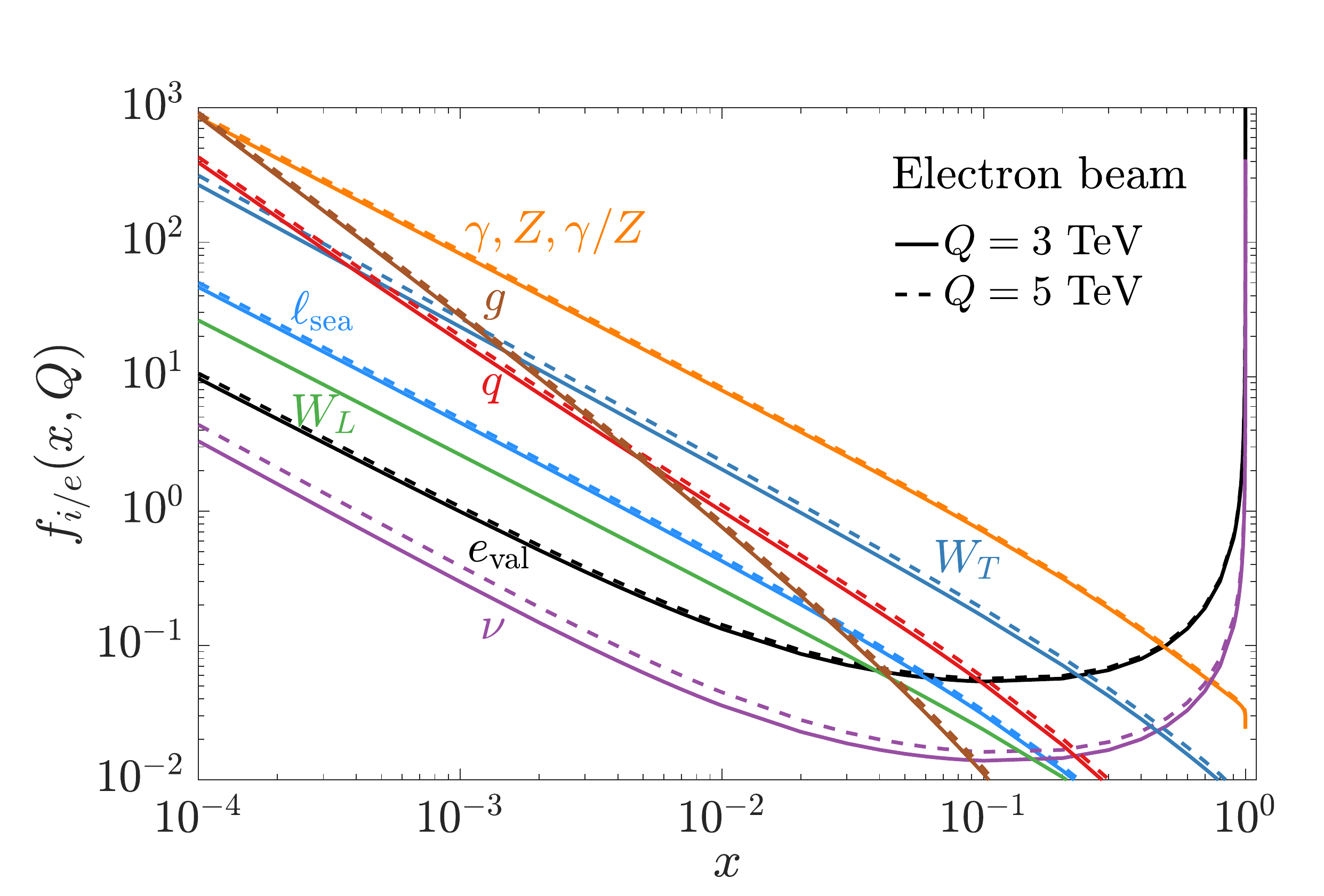}
\includegraphics[width=.48\textwidth]{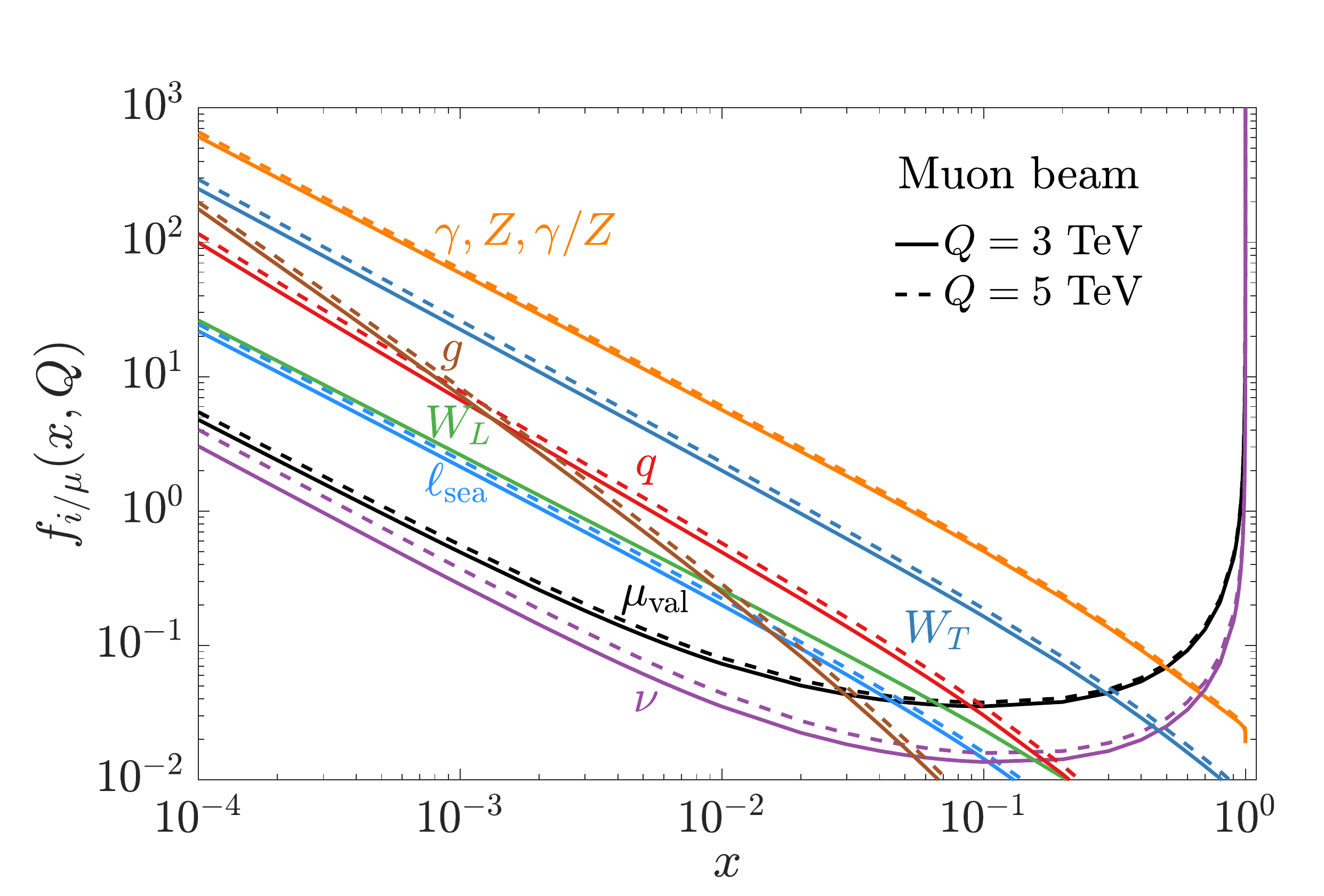}
\caption{PDFs in a high-energy lepton for (a) an electron and (b) a muon below the EW scale 
at $Q=30\ (50)~\gev$; and for (c) an electron and (d) an muon above the EW scale at $Q=3\ (5)~\tev$. }
\label{fig:PDFs}
\end{figure}

At the low energy below $\muEW$, the massive gauge bosons, neutrinos, and the top quark are inactive. We only have the PDFs for the flavors specified in Eq.~(\ref{eq:nf}) plus the photon and gluon.
We show the PDFs for 
an electron beam ($e^\pm$) in Fig.~\ref{fig:PDFs}(a) and 
a muon beam ($\mu^\pm$) in Fig.~\ref{fig:PDFs}(b) for the factorization scales $Q=30~(50)$ GeV. 

The initial condition for a valence lepton PDF is set as in Eq.~(\ref{eq:init}).
Including the leading soft radiation near $x\to1$, it behaves as $1/(1-x)$.
In the low-$x$ limit ($x\to0$), the valence PDF deviates from the leading $1/(1-x)$ behavior, and receives $1/x$ (and $\log x$) enhancement from higher order splitting $\gamma\to \ell^+\ell^-$. It coincides with sea flavor $f_{\lval}\sim f_{\bar{\ell}_{\textrm{val}}}$ shown explicitly in Fig.~\ref{fig:PDFs}, because $\gamma\to \ell^+\ell^-$ splitting gives the same amount of $\ell^+$ and $\ell^-$.

The photon is generated dynamically through the splitting of charged particles, $\ell(q)\to \ell(q)\gamma$. The leading order splitting gives the traditional Equivalent Photon Approximation (EPA) \cite{vonWeizsacker:1934nji,Williams:1934ad} 
\begin{equation}
\label{eq:EPA}
f_{\gamma/\ell,\textrm{EPA}}(x_\gamma,Q^2)=\frac{\alpha}{2\pi}\frac{1+(1-x_\gamma)^2}{x_\gamma}\log{Q^2 \over m_\ell^2} , 
\end{equation}
with a suitably chosen scale $Q$ associated with the physical process.\footnote{For consistency of the evolution and simplicity, we have only kept the leading-log term for the photon splitting. The non-log term corrections \cite{Budnev:1974de,Frixione:1993yw} may be sizable and become relatively more relevant for a muon collider.}
All the sea fermions, including leptons and quarks, are generated through $\gamma\to\ell^+\ell^-,q\bar{q}$, while gluon comes from $q\to qg$ splitting. In the low-$x$ limit, the generated PDFs behave as $1/x$ plus $\log^p x$ corrections. 

Including higher orders, the valence PDF receives threshold corrections of the form $1/(1-x)$ and $\log^p(1-x)$.
The precise determination of the PDFs in the $x\to1$ limit requires all orders of resummation. 
It can be only achieved for the valence non-singlet PDF under the fixed coupling assumption when $x$ asymptotically approaches 1, as demonstrated in App.~\ref{app:largeX}. Determination of the PDFs at other nontrivial $x$ value ($0<x<1$) or with a running coupling requires fully solving the DGLAP equations numerically. We outline the techniques we develop and take the non-singlet PDF of valence lepton as an example for demonstration in App.~\ref{app:solution}, while leave the comprehensive details of singlet, photon and gluon PDFs for a future work \cite{ours}.
A smooth transition to the $x\to1$ asymptotic form requires a consistent matching~\cite{Bertone:2019hks}.
In our practical treatment, we take the valence lepton PDF as a functional form as
\begin{equation}\label{eq:reg}
f_{\ell/\ell}(x,Q^2)=\begin{cases}
f_{\rm resum}(x,Q^2), & x < 1-\epsilon,\\
L(Q^2)\delta(1-x) ,   & x \geq 1-\epsilon,
\end{cases}
\end{equation}
where $\epsilon$ serves as a regulator.\footnote{Below the EW scale, 
we take $\epsilon =10^{-6}$. For EW PDFs above $\muEW$, we apply a more severe truncation $\epsilon=M_Z/Q$ to assure the correct double-log behavior in the $f\to fZ(f'W)$ splitting \cite{Bauer:2017isx,Han:2020uid}.}  
Within $x < 1-\epsilon$, the $f_{\rm resum}(x,Q^2)$ is obtained through the DGLAP resummation, which will converge to the all-order resummation form with a sufficient higher order of iterations, demonstrated in App.~\ref{app:solution}. Beyond the cutoff, the dynamically generated PDFs are negligible, while the valence PDF is taken as the form of a local form, $L(Q^2)\delta(1-x)$. The coefficient $L(Q^2)$ is determined through the momentum conservation \cite{Bauer:2017isx,Bauer:2018arx},
 \begin{equation}\label{eq:momavg}
\sum_i \langle x_i\rangle=1, ~\textrm{where}~ \langle x_i\rangle=\int xf_i(x,Q^2)\dd x.
\end{equation}
The index $i$ runs through all the flavors, including the leptons, photon, light quarks, and gluon below $\muEW$, as well as neutrinos, weak gauge bosons $W^{\pm}/Z$ and top quark above $\muEW$. The momentum conservation in Eq. (\ref{eq:momavg}) ensures a cancellation of the regulator $\epsilon$ between the local term $L(Q^2)$ and the integration over $x<1-\epsilon$ in a physical observable computation. 

As discussed in Sec.~\ref{sec:QCD}, degeneracies exist for the sea leptons, up-type and down-type quarks as in Eqs.~(\ref{eq:Lsea}-\ref{eq:Dsea}). The leading splittings $\gamma\to \ell^+\ell^-,q\bar{q}$ result in the approximate ratio for one flavor in the moderate $x$ region
\begin{equation}
\label{eq:Rsea}
f_{\bar{\ell}_{\textrm{val}}}:f_{u}:f_{d}\sim 1:N_ce_u^2:N_ce_d^2=1:\frac{4}{3}:\frac{1}{3},
\end{equation}
where $N_c=3$.
At small $x$, the light-quark ($u$- and $d$-type) PDFs merge due to the resummation of large and universal QCD logarithmic terms ($\alpha_s\log x$). In the relatively large $x$ region ($x\gtrsim0.5)$, the energetic quarks tend to radiate more than leptons and $f_u$ even becomes slightly smaller than $f_{\bar{e}}$, as a result of the additional QCD splitting $q\to qg$.
For a muon beam ($\mu^\pm$), 
$\log(Q^2/m_e^2)/\log(Q^2/m_\mu^2) \sim 2$ at $Q\sim30\ (50)$ GeV. The QCD partons (quark and gluon) in the electron beam are significantly larger than those in the muon beam, because of the accumulation of the large QCD log terms. We also note that the PDF uncertainties due to the scale choices of 30 GeV and 50 GeV are moderate, about 10\% for $f_{g/e}$ and $20\%$ for $f_{g/\mu}$. Besides, we have also estimated the QCD threshold uncertainty by varying the matching scale as $\muQCD=0.7$ GeV \cite{Drees:1994eu}, which is less than 20\% (10\%) for an electron (muon) beam~\cite{Buarque:2021dji}.

It is informative to consider the PDF evolution above the EW scale. We thus also show the full EW PDFs at high scales of 3 (5) TeV in Figs.~\ref{fig:PDFs}(c) and (d). In these plots, we have summed over the non-valence fermions as
\begin{equation}
f_{\lsea}=f_{\bar{\ell}_{\textrm{val}}}+\sum_{i\neq\lval}^{N_{\ell}} (f_{\ell_i}+f_{\bar{\ell}_i}), \ \ 
f_{\nu}=\sum_{i}^{N_{\ell}} (f_{\nu_i}+f_{\bar{\nu}_i} ), \ \ 
f_{q}=\sum_{i}^{N_u}(f_{u_i}+f_{\bar{u}_i})+\sum_{i}^{N_d}(f_{d_i}+f_{\bar{d}_i}) .
\label{eq:sea}
\end{equation}
Here, $N_u=3$ as the top quark becomes active as well.
The neutral-current EW PDFs include $\gamma$, $Z$, and $\gamma Z$-mixing. 
The longitudinal PDFs $(W_L,Z_L)$ were known at the leading order as the Effective $W$ Approximation \cite{Kane:1984bb,Dawson:1984gx,Chanowitz:1985hj}, 
which do not run with the scale $Q$, as an explicit realization of the Bjorken-scaling restoration. 
We find that the EW corrections from $W/Z$ to the light particle PDFs at a high scale above TeV can be as large as 50\% (100\%) for $f_{d/e}~(f_{d/\mu})$, due to the relatively large SU(2)$_{\rm L}$ gauge coupling compared with the electromagnetic one. The scale choices of 3 TeV and 5 TeV give uncertainty about 15\% (20\%) in the electron (muon) beam.
The detailed comparison and potential physical impacts are left for a future publication \cite{ours}.

It is interesting to ask how much momentum each parton species carries along the longitudinal beam direction. We explicitly show the average momentum fractions $\langle x_i\rangle$ carried by a parton $i$ in Table \ref{tab:momavg}. Our results are shown for both an electron beam in (a) and a muon beam in (b). 
Naively, the momentum ratio for the sea leptons and quarks may be estimated by Eq.~(\ref{eq:Rsea}) as 
\begin{equation}
\frac{\langle x_q\rangle}{\langle x_{\lsea}\rangle}\lesssim
\frac{N_c\left[\sum_i(e_{u_i}^2+e_{\bar{u}_i}^2)+\sum_{i}(e_{d_i}^2+e_{\bar{d}_i}^2)\right]}
{e^2_{\bar{\ell}_\textrm{val}}+\sum_{i\neq\lval}(e_{\ell_i}^2+e_{\bar{\ell}_i}^2)}
=\frac{22/3}{5}.
\end{equation}
The actual numbers in Table \ref{tab:momavg} are smaller than this estimation, as pointed out that gluon takes part of the quark momentum fractions. 
After adding the gluon contribution, we obtain an improved estimation 
\begin{equation}
\frac{\langle x_q\rangle+\langle x_g\rangle}{\langle x_{\lsea}\rangle}\simeq\frac{22/3}{5}.
\end{equation}
Table \ref{tab:momavg} gives us the relative size of each parton species and the variation at a few representative scales. In addition, we see that there is less radiation and thus less sea quark contribution for a muon beam than an electron beam. 

\begin{table}[tb]
\begin{tabular}{|c|c|c|c|c|c|}
\hline
$Q(e^\pm)$ & $e_{\textrm{val}}^{}$ & $\gamma$ & $\lsea$ & $q$ & $g$\\
\hline
30 GeV &96.6 & 3.20 & 0.069 & 0.080 & 0.023 \\
50 GeV &96.5 & 3.34 & 0.077 & 0.087 & 0.026 \\
$M_Z$ &96.3 & 3.51 & 0.085 & 0.097 & 0.028 \\
\hline
\end{tabular}
\begin{tabular}{|c|c|c|c|c|c|}
	\hline
	$Q(\mu^\pm)$ & $\mu_{\textrm{val}}^{}$ & $\gamma$ & $\lsea$ & $q$ & $g$\\
	\hline
	30 GeV & 98.2 & 1.72 & 0.019 & 0.024 & 0.0043 \\
	50 GeV & 98.0 & 1.87 & 0.023 & 0.029 & 0.0051\\
	$M_Z$ & 97.9 & 2.06 & 0.028 & 0.035 & 0.0062 \\
	\hline
\end{tabular}
\caption{The averaged momentum fractions [\%] carried by each parton species for (a) an electron beam and (b) a muon beam with a few representative values of the factorization scale $Q$.} 
\label{tab:momavg}
\end{table}

\begin{figure}\centering
\includegraphics[width=.48\textwidth]{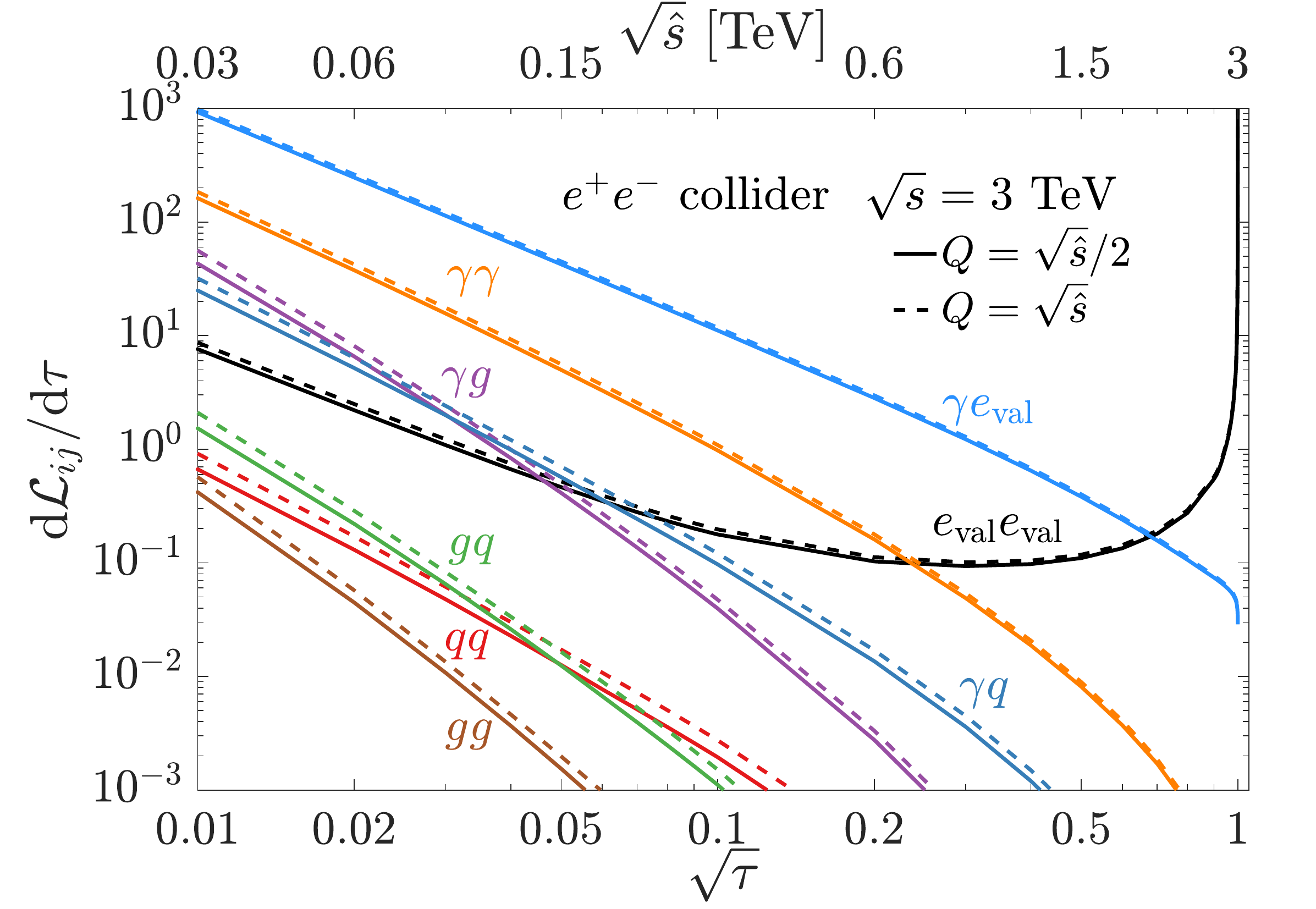}	
\includegraphics[width=.48\textwidth]{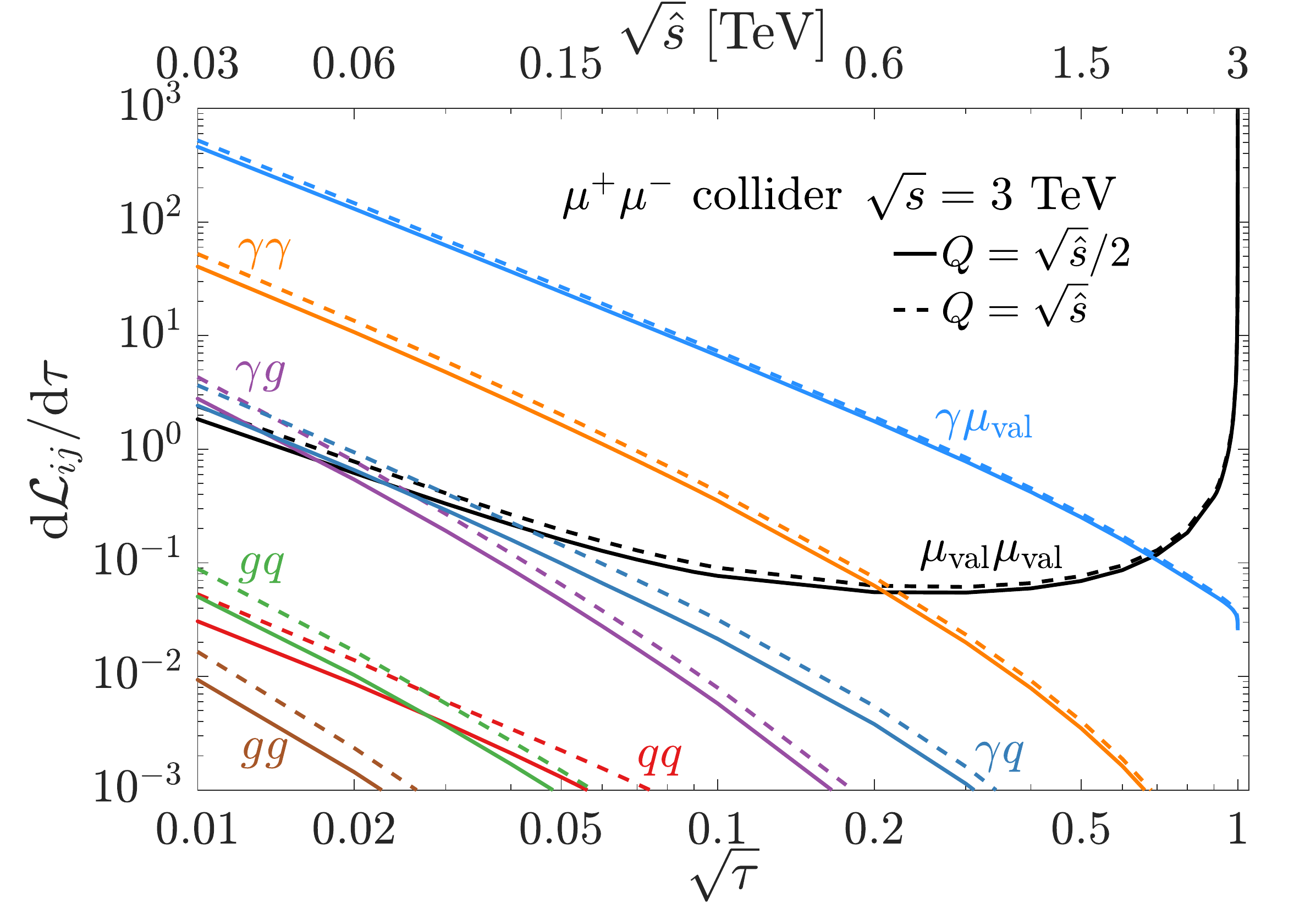}	
\includegraphics[width=.48\textwidth]{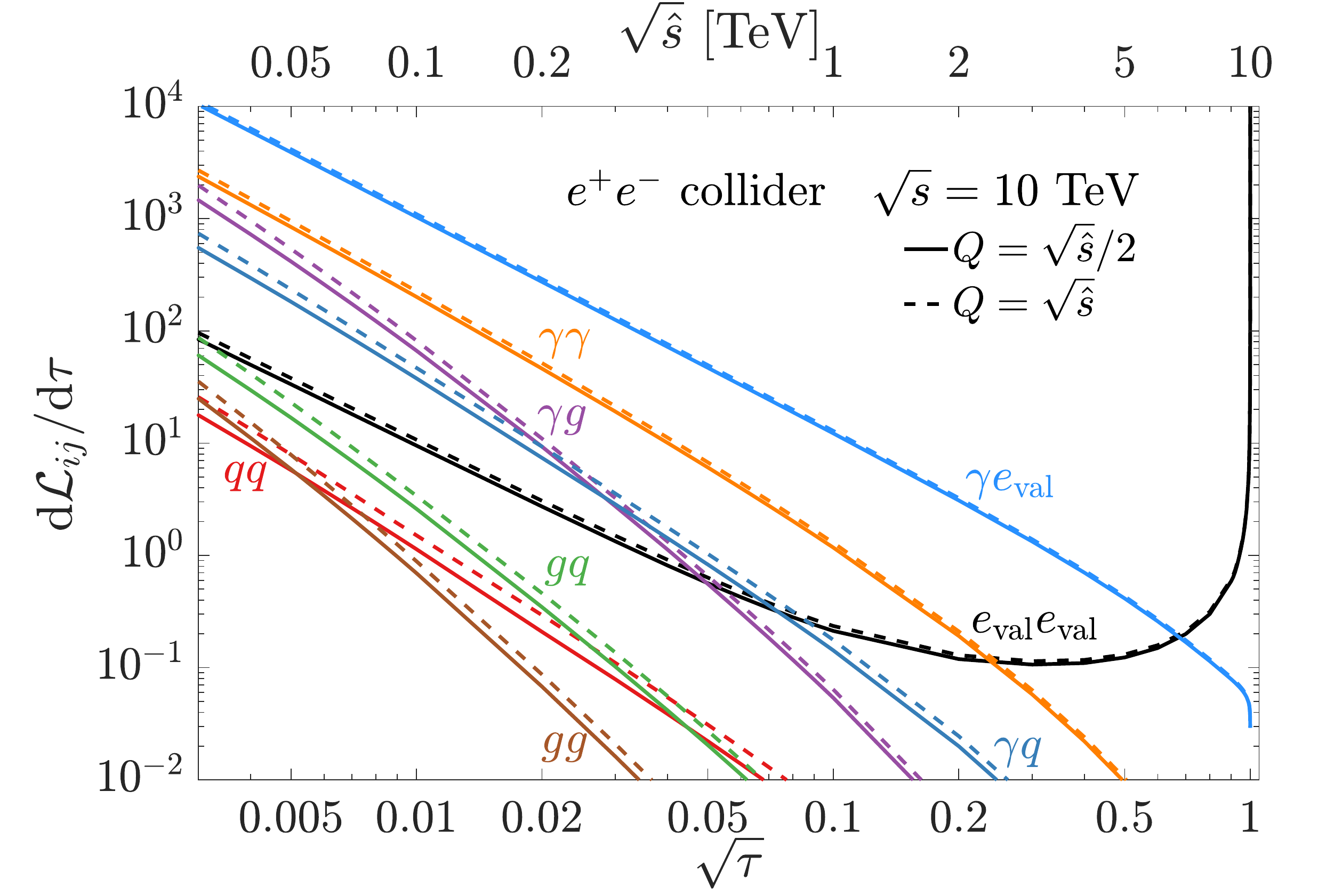}	
\includegraphics[width=.48\textwidth]{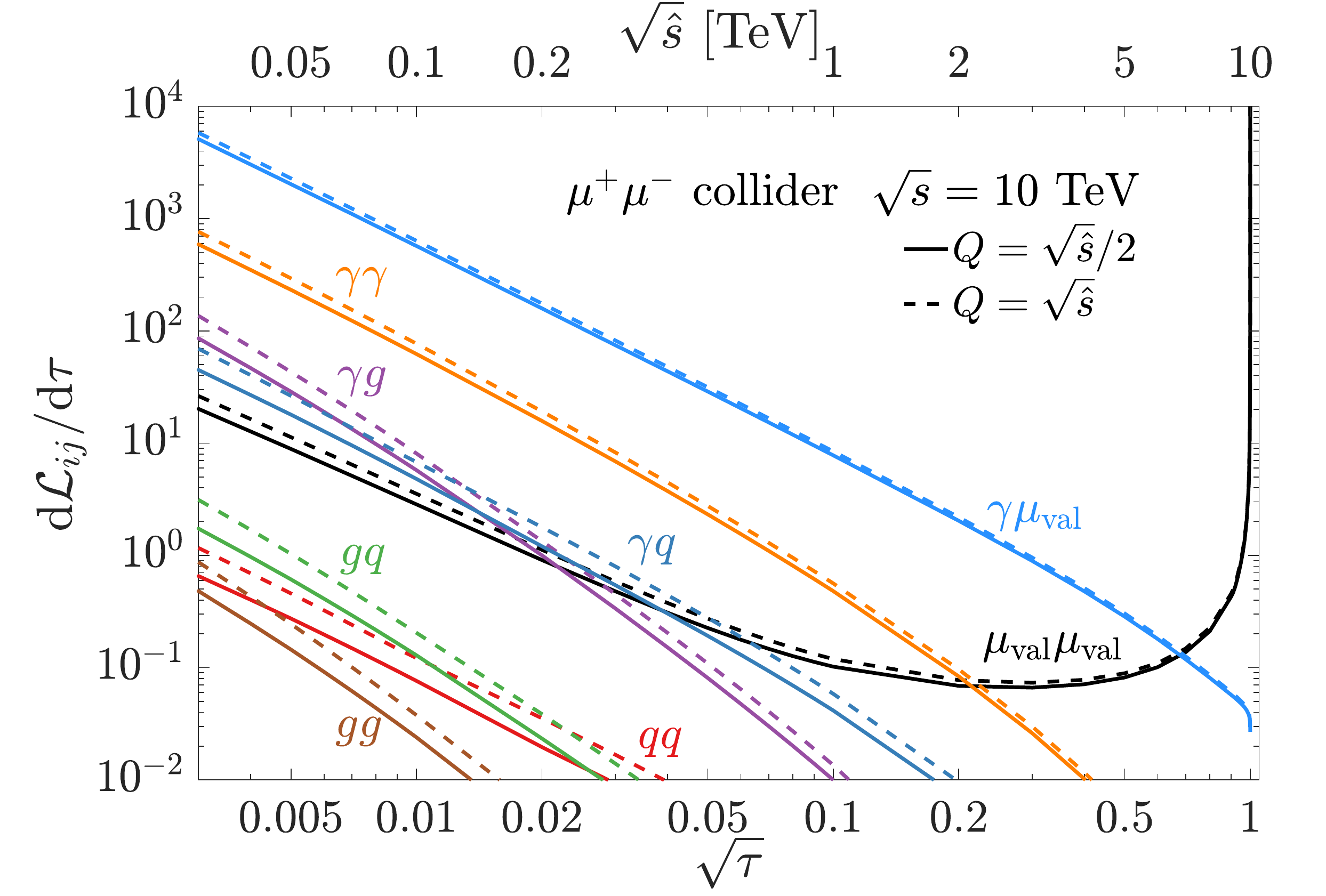}	
\caption{Parton luminosities $\dd\mathcal{L}_{ij}/\dd\tau$ for 
(a) an $e^+ e^-$ collider at $\sqrt s = 3$ TeV, 
(b) a $\mu^+\mu^-$ collider at $\sqrt s = 3$ TeV,
(c) an $e^+ e^-$ collider at $\sqrt s = 10$ TeV, 
and (d) a $\mu^+\mu^-$ collider at $\sqrt s = 10$ TeV.
The factorization scale is chosen as $Q=\sqrt{\hat s}/2$ (solid curves) and $\sqrt{\hat s}$ (dashed curves).
}
\label{fig:lumi}
\end{figure}

To make the connection with the physical scattering processes, we next compute the partonic luminosities for the initial states 
\beq 
\ell^+\ell^-,\ \gamma \ell,\ \gamma \gamma,\ qq,\ \gamma q, ~\gamma g,\ gq\ {\rm and}\ gg, 
\label{eq:partons}
\eeq
for $\sqrt s = 3$ TeV and 10 TeV, as shown in Fig.~\ref{fig:lumi} versus $\sqrt{\tau} = \sqrt{\hat s/s}$, the ratio of the partonic c.m.~energy and the collider energy,
where the sea fermion species are summed as in Eq.~(\ref{eq:sea}).
We see that a high-energy lepton collider can offer a broad spectrum of initial state particles.  
Of our particular interests, 
the QCD parton luminosities involving quarks and gluons increase significantly at low 
$\sqrt{\tau}$. The parton luminosities of $\gamma g+\gamma q$ are about $50\%\ (20\%)$ of that of $\gamma\gamma$ for an $\ee$ ($\mm$) collider. 
The QCD parton luminosities of $qq, gq$ and $gg$ are about $2\%\ (0.5\%)$ of that of $\gamma\gamma$ 
for an $\ee$ ($\mm$) collider. 
Correspondingly, given the stronger coupling over QED, we may expect sizable QCD cross sections at low $\sqrt{\tau}$. 
Our standard choice for the factorization scale is
\beq\label{eq:scale}
Q=\sqrt{\hat{s}}/2.
\eeq
Varying the scale from this default choice (solid curves) to $Q=\sqrt{\hat s}$ may result in a luminosity uncertainty of $20\%\ (50\%)$ for a photon-initiated (gluon initiated) process. 

\section{The standard processes and jet production}
\label{sec:SMProcesses}

\subsection{EW processes}
\label{sec:EW}

In high-energy $\ee$ collisions, one would expect that the leading reactions are of the QED and electroweak nature, including Bhabha scattering $e^+e^-\to e^+e^-$, Compton scattering $\gamma e \to \gamma e$, and the $s$-channel annihilation processes for pair production $\ee \to \mm, q\bar q$ and $W^+W^-$ once above the threshold. While the cross sections for the annihilation processes fall with the c.m.~energy as $\sigma \sim \alpha^2/s$, the $t$-channel processes receive the collinear enhancement. Nevertheless, with a detector angular acceptance $\theta_{\min}$, the cross sections for the $2\to 2$ $t$-channel processes still fall as $\sigma \sim \alpha^2 /(s\, \theta_{\min}^2)$. Going beyond the fixed-order calculations, the potentially large collinear logarithms ($\log\theta^2$) need to be resummed, leading to the appropriate description of the parton distribution functions (PDFs), as presented in the previous section. As such, there will be substantial contributions coming from partonic scattering processes initiated by those in Eq.~(\ref{eq:partons}), far below the collider c.m.~energy. 
Throughout this work, the partonic cross sections are calculated at the leading order with the general purpose event generator \textsc{MadGraph5} v2.6.7 \cite{Alwall:2014hca}. The annihilation processes with the initial-state radiation (ISR) are calculated with \textsc{Whizard} v2.8.5 \cite{Kilian:2007gr}. 
%

\begin{figure}[tb]
\centering
\includegraphics[width=.48\textwidth]{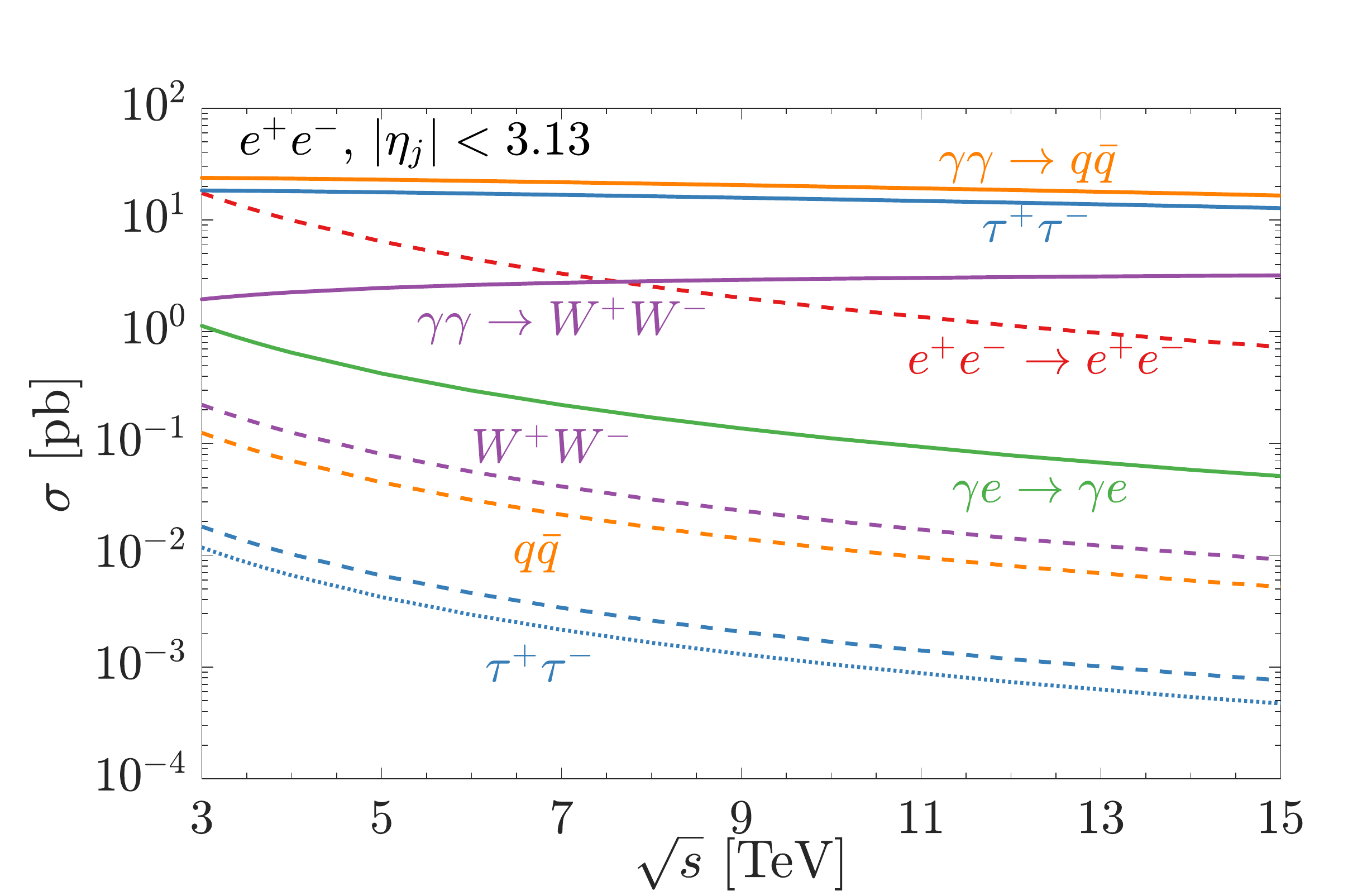}
\includegraphics[width=.48\textwidth]{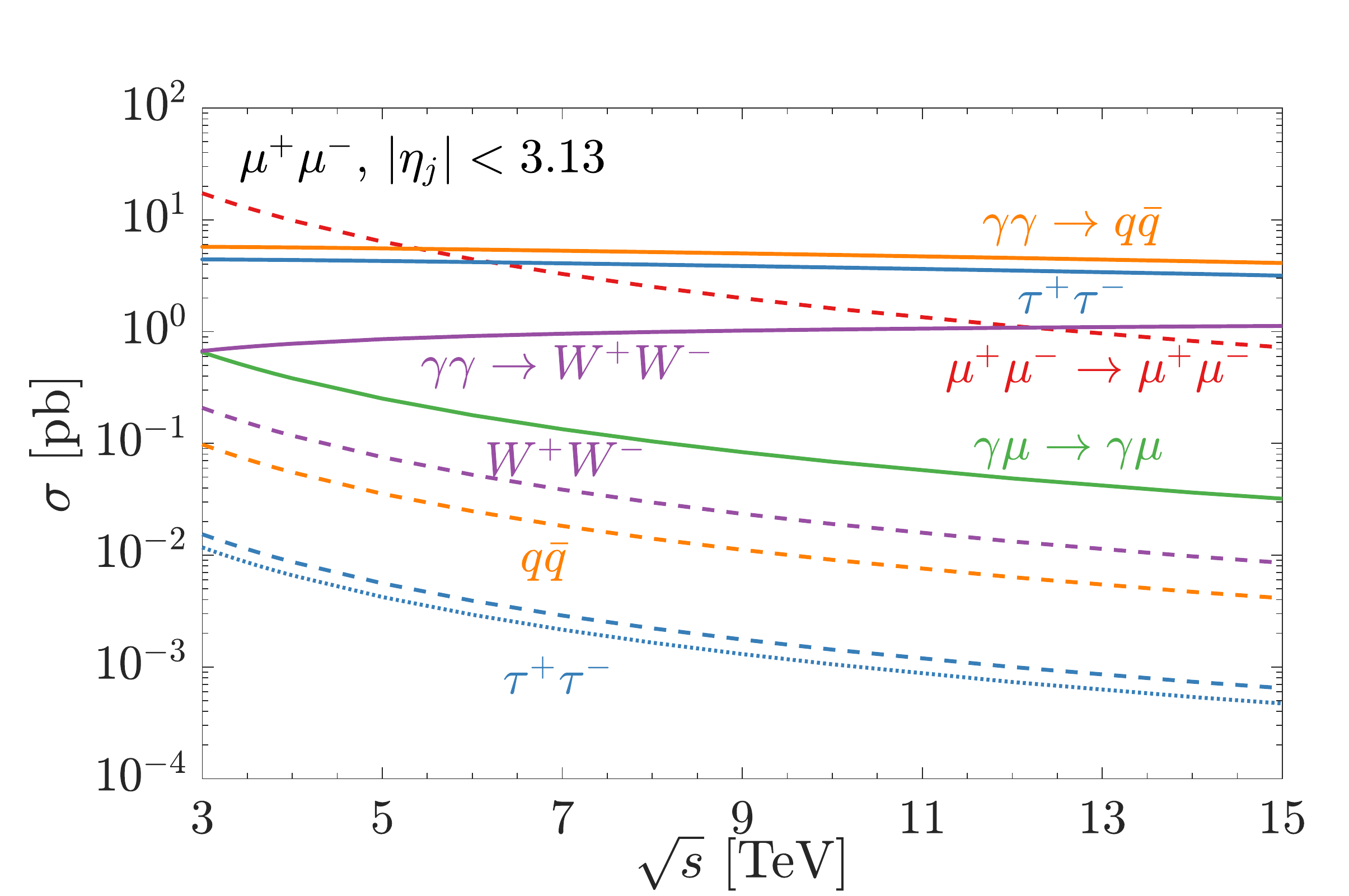}
\includegraphics[width=.48\textwidth]{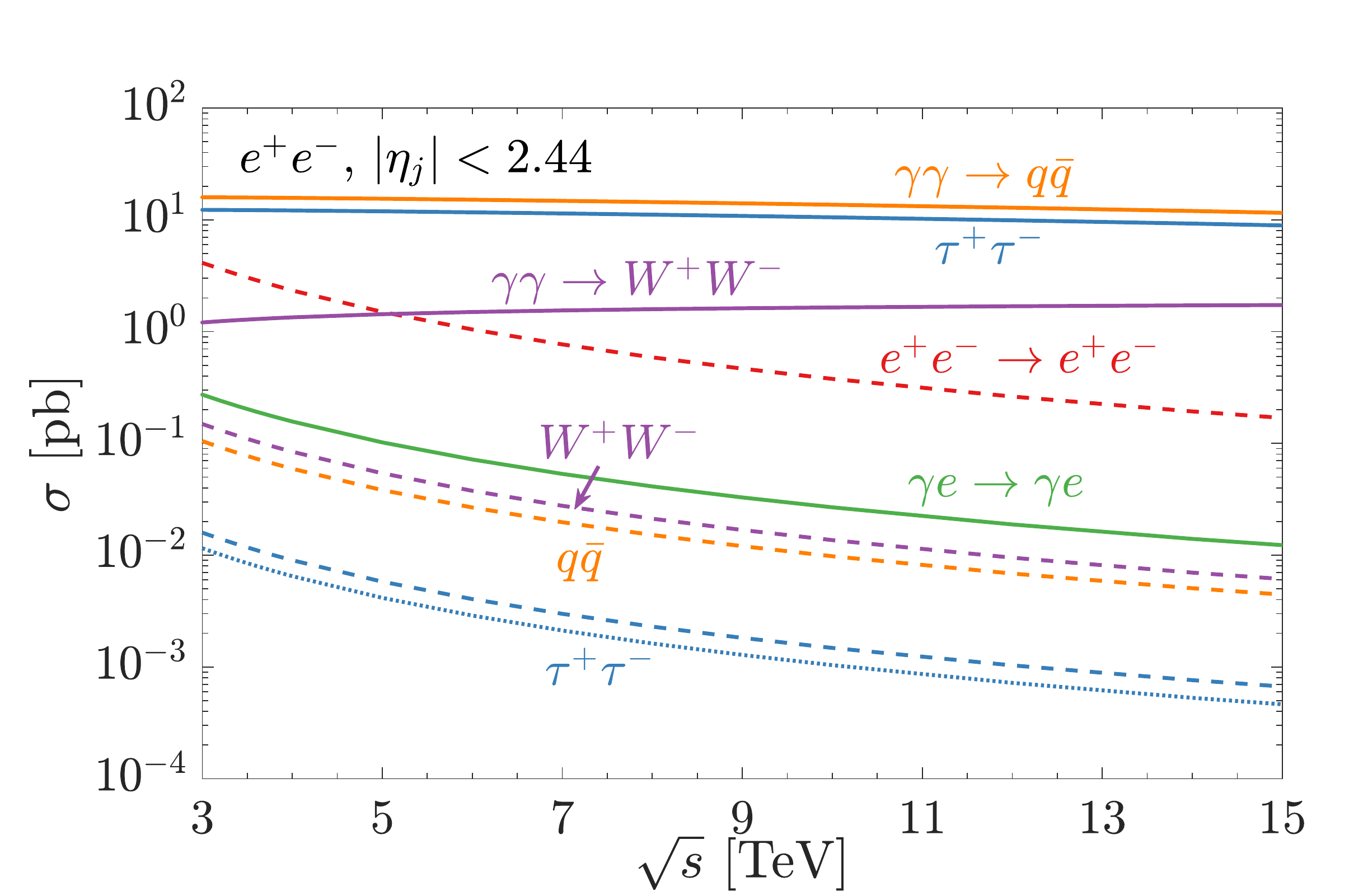}
\includegraphics[width=.48\textwidth]{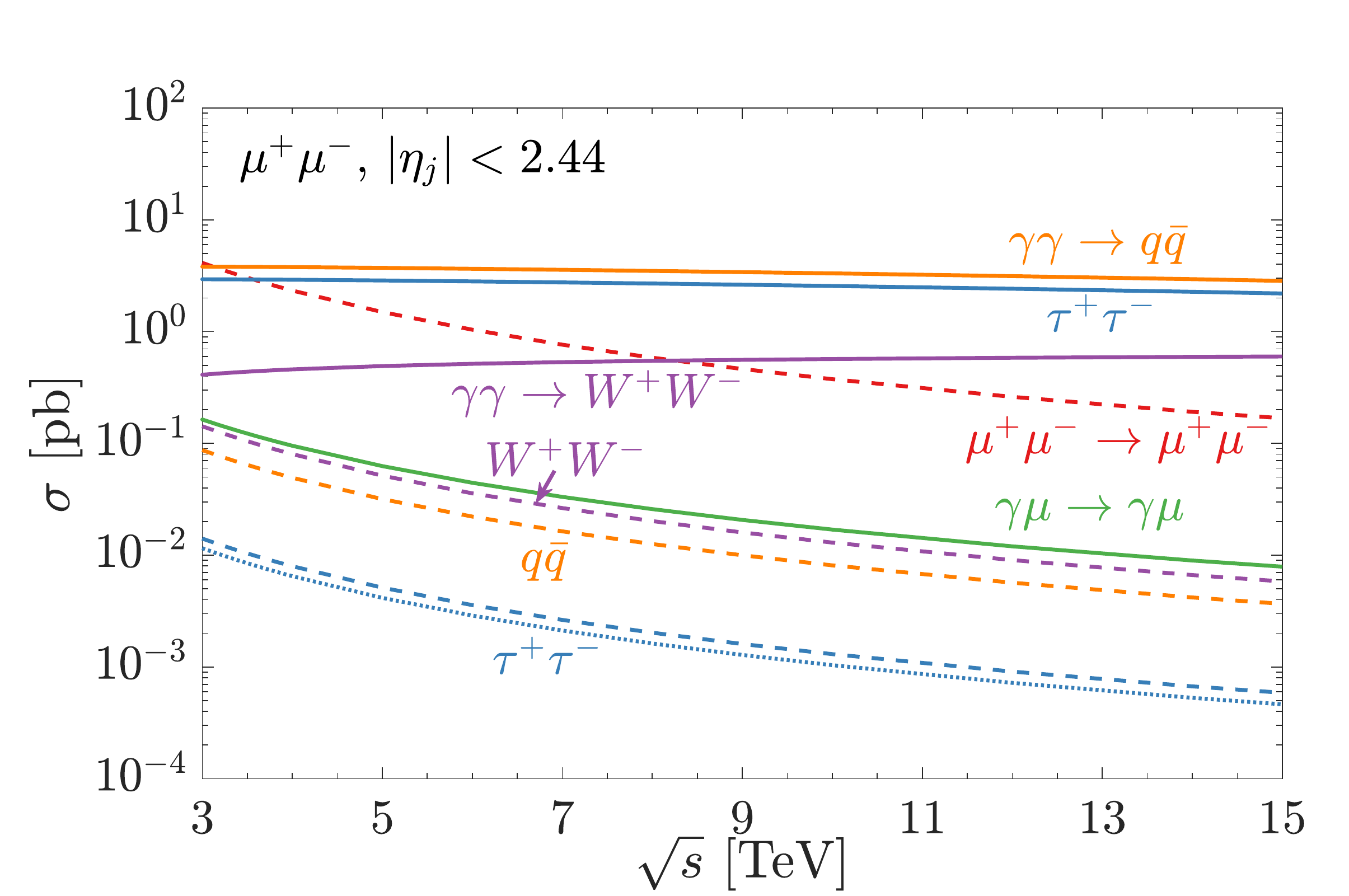}
\caption{Cross sections for the annihilation processes versus the collider c.m.~energy for an $\ee$ collider (left panels) and a $\mm$ collider (right panels) with basic acceptance cuts in Eq.~(\ref{eq:cut}). 
The downward dashed (dotted for $\tau^+\tau^-$) curves indicate the corresponding Bhabha scattering and $\ell^+\ell^-$ annihilation processes with (without) ISR.
}
\label{fig:SMXS}
\end{figure}

We first present some leading order production cross sections of typical electroweak processes in Fig.~\ref{fig:SMXS} versus the collider c.m.~energy for both an $\ee$ collider (left panels) and a $\mm$ collider (right panels), including the effects of ISR \cite{Greco:2016izi}. 
In Fig.~\ref{fig:SMXS}, the dashed (falling) curves represent the Bhabha scattering and annihilation processes
\beq
\ell^+\ell^- \to \ell^+\ell^-,\ \tau^{+}\tau^{-},\ q\bar{q}\ {\rm and}\ W^+W^-. 
\label{eq:anni}
\eeq
The cross sections scale as $1/s$, with the characteristic kinematics of the final-state pair invariant mass close to the collider energy $m_{ij}\approx \sqrt s$. At high energies, the ISR effects reduce the effective partonic collision energy $\hat{s}$ and thus increase the cross sections $\sim 1/\hat{s}$. For illustration, we compare the result without ISR for $\ell^+\ell^- \to \tau^{+}\tau^{-}$ by the dotted curves in the panels. Typically, the effective reduction is about a factor of 20\%$-$80\% (10\%$-$40\%) for an electron (muon) collider. The radiative returns to the $Z$ resonant production also enhance the light-particle cross sections significantly. 
The ISR effects for light-particle production ($\tau^+\tau^-$, $q\bar{q}$) are thus larger than the massive one ($W^+W^-$), because of the lower threshold, \emph{i.e.}, $\hat{s}>m_{ij}^2$ versus $\hat{s}>(2 M_W)^2$. 

In considering the QED fusion processes, the initial state partons present an infrared enhancement at low $m_{ij}$ and the two-parton cross section scales as 
\beq
\sigma \sim {\alpha^2 \over m_{ij}^2} \left({\alpha \over 2\pi} \log{Q^2 \over { m_{\ell}^2} }\right)^2.
\label{eq:xSscale}
\eeq
To separate the hadronic activities with the low-momentum transfer from the hard processes of our current interests, we impose the following basic acceptance cuts on the outgoing particles in the transverse momentum 
$(p_T^j)$, the di-jet invariant mass and the pseudo-rapidity $(\eta_j)$ in the lab frame 
\begin{equation}
\label{eq:cut}
p_T^j > \left(4 + {\sqrt s \over 3\ {\rm TeV}}\right) \GeV, \quad m_{ij}> 20 \GeV, \quad |\eta_j|<3.13\ (2.44).
\end{equation}
The energy-dependent cut on the final state $p_T^j$ is to uniformly control the collinear logs of the form 
{$(\alpha_s/\pi) \log(p_T^j/\sqrt s)$}, numerically motivated by a CLIC study \cite{Barklow:2011aa}. The pseudo-rapidity cut corresponds to an angle with respect to the beam in the lab frame $\theta_j\sim5\degree\ (10\degree)$, in accordance with the detector coverage. For an equal footing comparison, the same acceptance cuts have been applied to the Bhabha scattering and annihilation processes in Fig.~\ref{fig:SMXS} as well.

In Fig.~\ref{fig:SMXS}, the solid lines show the Compton scattering and the fusion processes 
\beq
\gamma\ell\to \gamma\ell ; \quad 
\gamma\gamma \to \ell^+\ell^-,\ q\bar{q}\ (u,d,c,s,b),\ {\rm and} \ W^+W^- ,
\eeq
by exploiting the EPA in Eq.~(\ref{eq:EPA}). 
The upper panels and lower panels are with a different rapidity (angle) cut as in Eq.~(\ref{eq:cut}). 
The cross section for the Compton scattering ($\gamma\ell$) also falls as $\alpha^2/(s\, \theta^2)$, as evidenced from the figures. 
The cross sections for the other fusion processes increase with energy logarithmically and decreases with $p_T$ (or $m_{ij}$) as in Eq.~(\ref{eq:xSscale}). The angular dependence is much weaker than $1/\theta^2$ and becomes roughly like $\eta^2$ due to the boost factor. 
We see that the fermion pair production can be larger than that of the $WW$ channel, which is known to be one of the leading channels for high-energy leptonic collisions. 
For the sake of illustration, we have only included the leading contributions from $\gamma\gamma$ fusion in Fig.~\ref{fig:SMXS}. We remind the reader that for the $W^+W^-$ production at these energies, the sub-leading channel $\gamma Z\to W^+W^-$ contributes to about 20\% (40\%), and $ZZ, W^+W^- \to W^+W^-$ about $10\%$ ($30\%$) concerning the $\gamma\gamma$ contribution at an $\ee$ ($\mm$) collider. They are neglected in our comparison for simplicity, which does not change the conclusion \cite{ours}.

\subsection{Jet production}
\label{sec:jets}

\begin{figure}
	\includegraphics[width=0.49\textwidth]{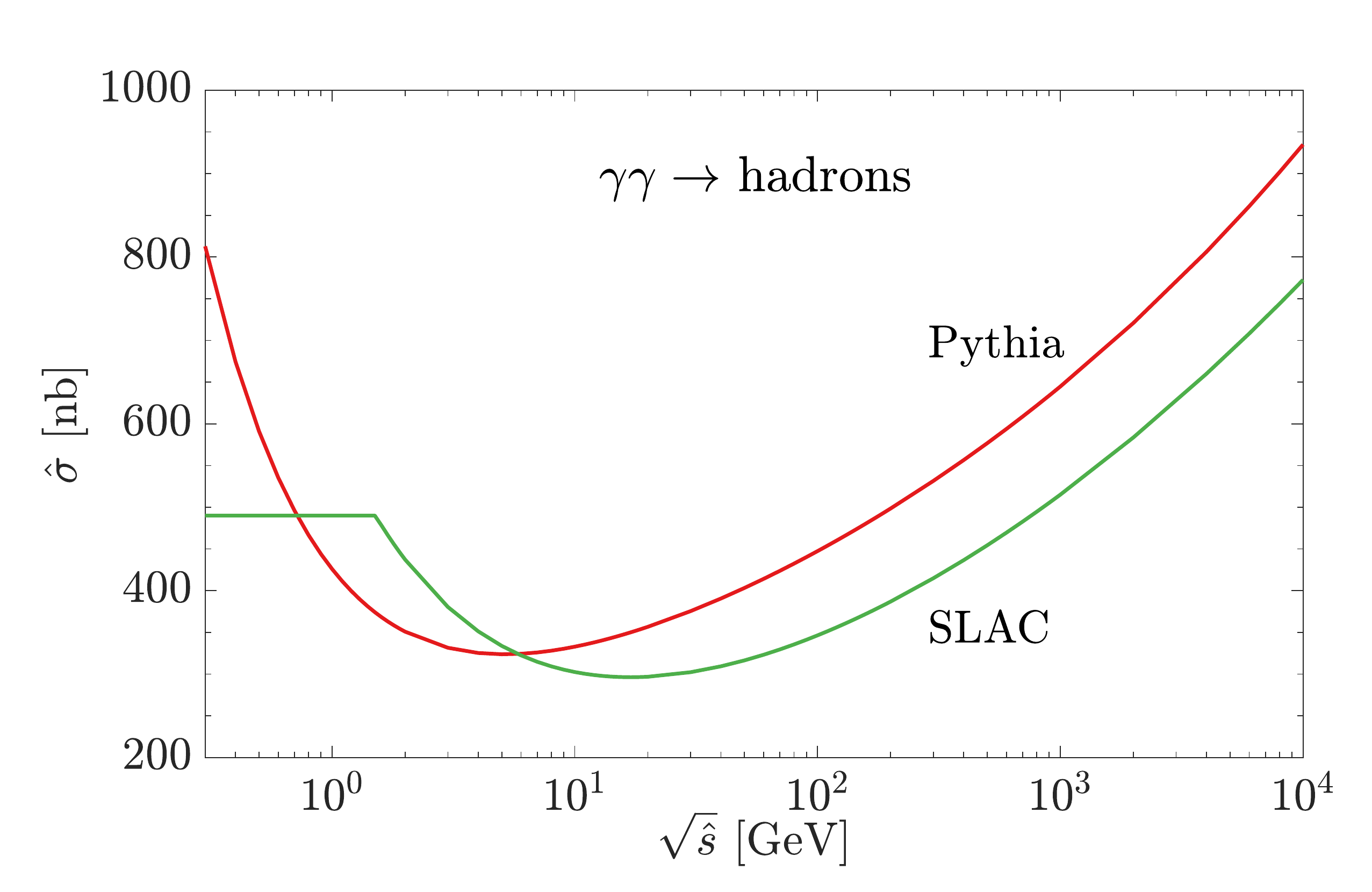}
	\includegraphics[width=0.49\textwidth]{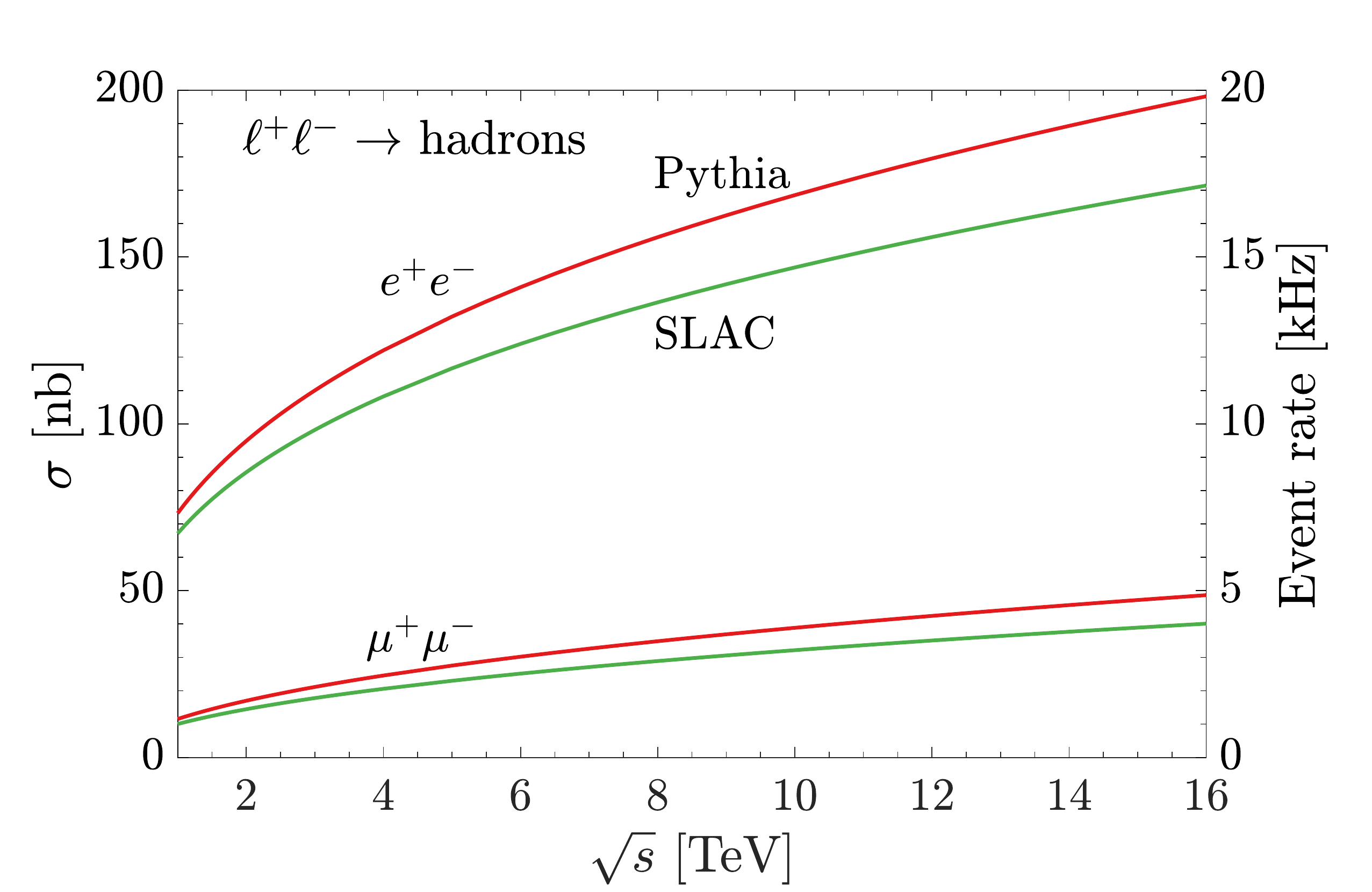}
	\caption{The photonic (a) and leptonic (b) cross sections for photon-induced hadronic production at high-energy lepton colliders. We adopted the models by Pythia~\cite{Schuler:1996en,Sjostrand:1993yb} or SLAC~\cite{Chen:1993dba} parameterizations as stated in the text.}
	\label{fig:hadron}
\end{figure}

Before predicting the jet production rate, it is important to remind the reader that at the low-momentum transfer, the majority of the events come from the hadronic production of the photon-induced processes, constituting the substantial backgrounds at the detector. This was pointed out in 
Refs.~\cite{Drees:1991zka,Drees:1992ws} for $\ee$ collisions in the context of beamstrahlung, and have been since extensively studied \cite{Chen:1993dba,Godbole:2011zz}. 
Similar to the behavior of the total cross sections in hadronic collisions \cite{Froissart:1961ux,Martin:1965jj}, the photon-induced hadronic cross section moderately increases with energy.
Due to the non-perturbative nature of the low-energy reactions, one would have to model the scattering. 
We estimate the total cross sections by adopting the two well-studied parameterizations for 
$\gamma\gamma\to\textrm{hadrons}$ in Pythia \cite{Schuler:1996en,Sjostrand:1993yb},
\begin{equation}
\hat\sigma_{\gamma\gamma}(\hat s) \approx (211~{\rm nb})~{\hat s}^{0.0808} + (215~{\rm nb})~{\hat s}^{-0.4525},
\end{equation}
 and by a SLAC group \cite{Chen:1993dba},
 \begin{equation}
 \hat\sigma_{\gamma\gamma}(\hat s) \approx
\begin{cases}
490~{\rm nb}&(0.3~{\rm GeV}< \sqrt{\hat s} < 1.5~{\rm GeV}),\\ 
200~{\rm nb}~[1+ 0.0063(\ln\hat s)^{2.1} + 1.96~{\hat s}^{-0.37}] &(\sqrt{\hat s}\geq 1.5~{\rm GeV}),
\end{cases}
\end{equation}
where $\hat s$ is the c.m.~energy squared for the $\gamma\gamma$ collisions in units of GeV$^2$.
We show the results for the photon-induced cross sections 
in Fig.~\ref{fig:hadron}. 
We see that the $\gamma\gamma$ cross section may reach the order of micro-barns ($\mu$b) at the TeV c.m.~energies. 
Folding in the $\gamma\gamma$ luminosity in electron/muon collisions,\footnote{Here we have neglected the effects of beamstrahlung. This is justifiable for the large muon mass and for the circular collider designs.} this brings the cross section down to the level of one hundred or a few tens of nano-barns at high-energy electron or muon colliders. 
The axis on the right indicates the event rate in kHz, assuming an instantaneous luminosity of $10^{35}/$cm$^2/$s.
Those hadronic final states dominate the event shape in this low energy regime. 
However, those events are typically populated at very small scattering angles and low transverse momenta below a few GeV \cite{Barklow:2011aa}. While they should be taken into account for the detector design and the experimentation, they would not have much impact on the high-$p_T$ physics of our current consideration.

\begin{figure}[tb]
	\centering
	\includegraphics[width=.48\textwidth]{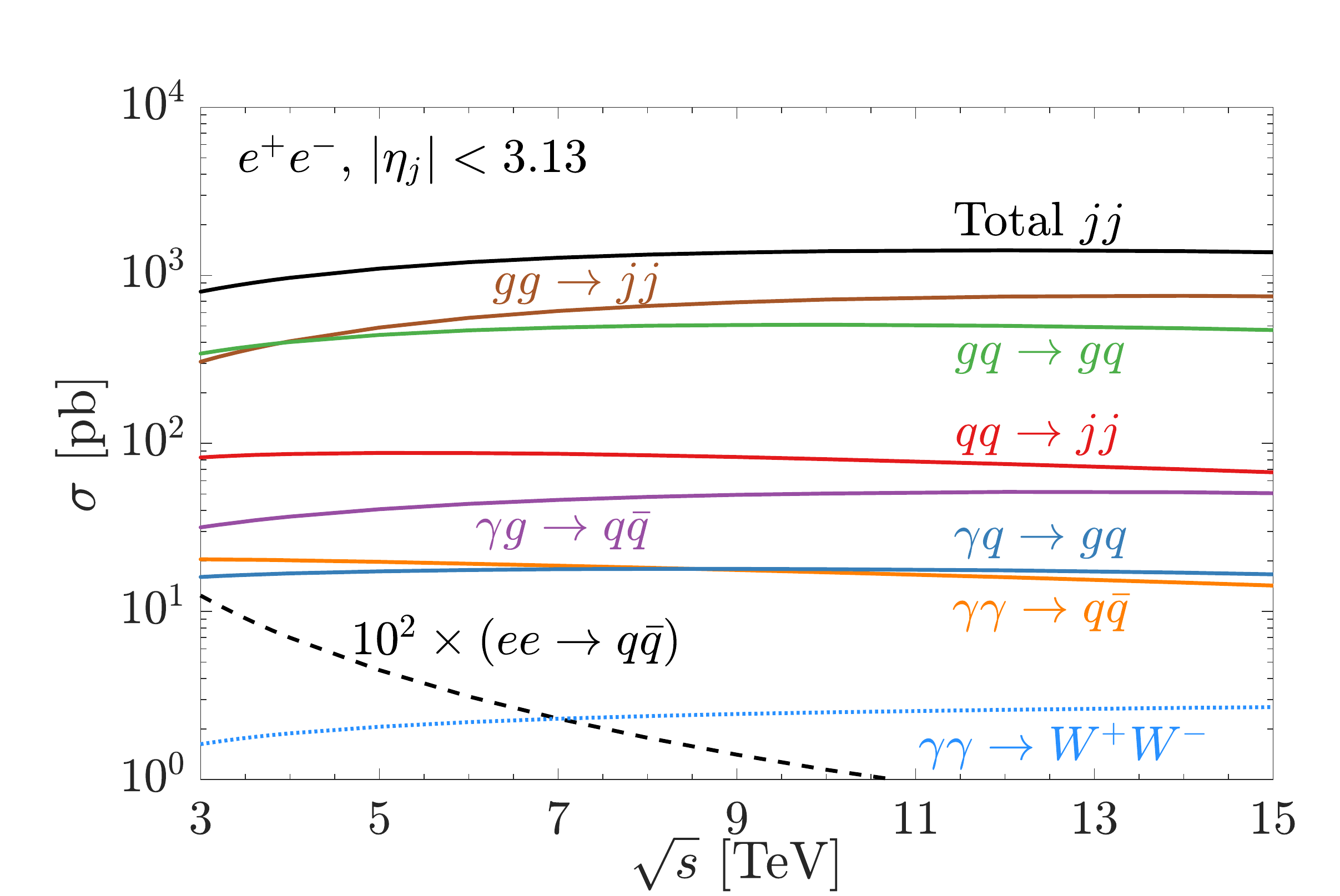}
	\includegraphics[width=.48\textwidth]{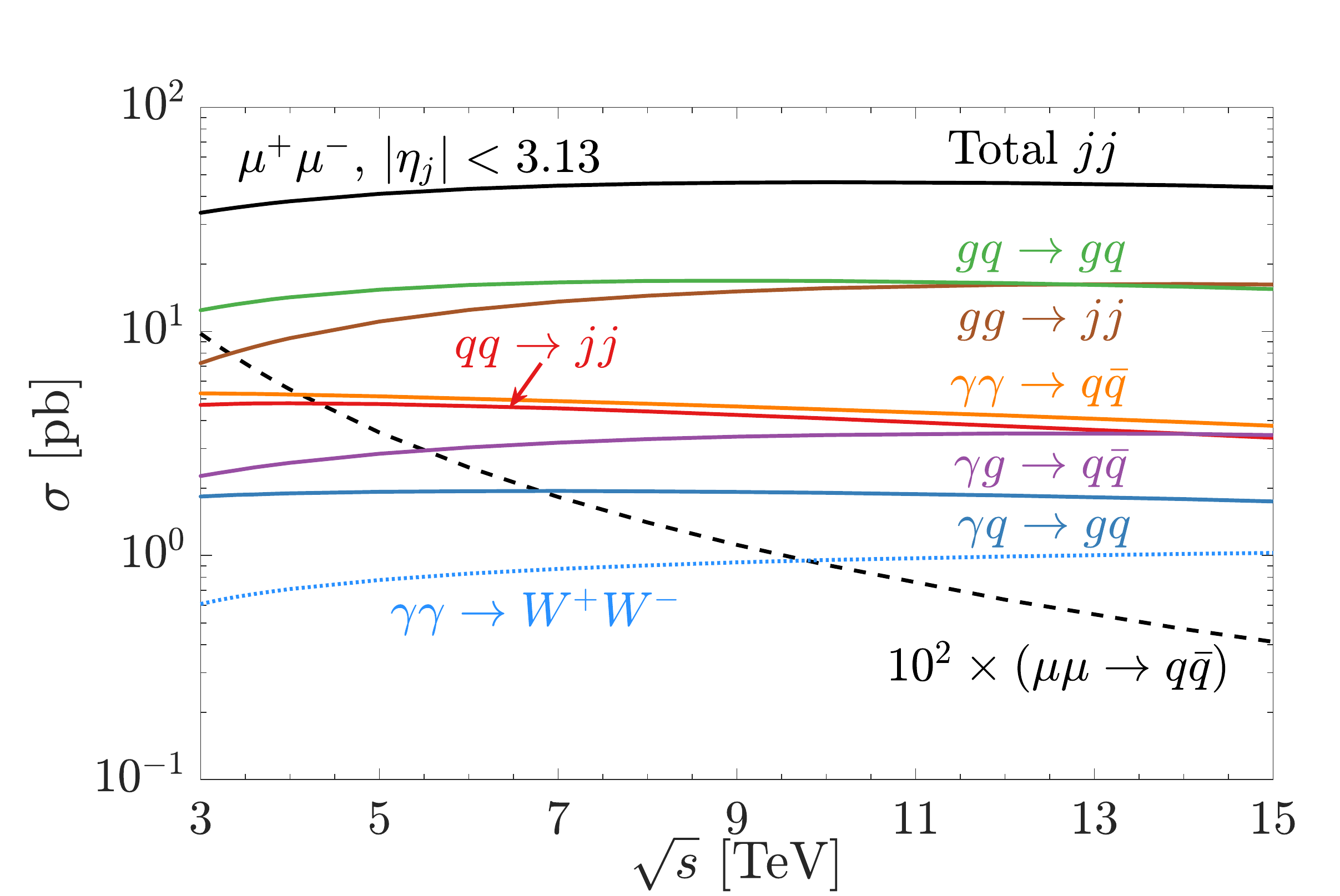}
	\includegraphics[width=.48\textwidth]{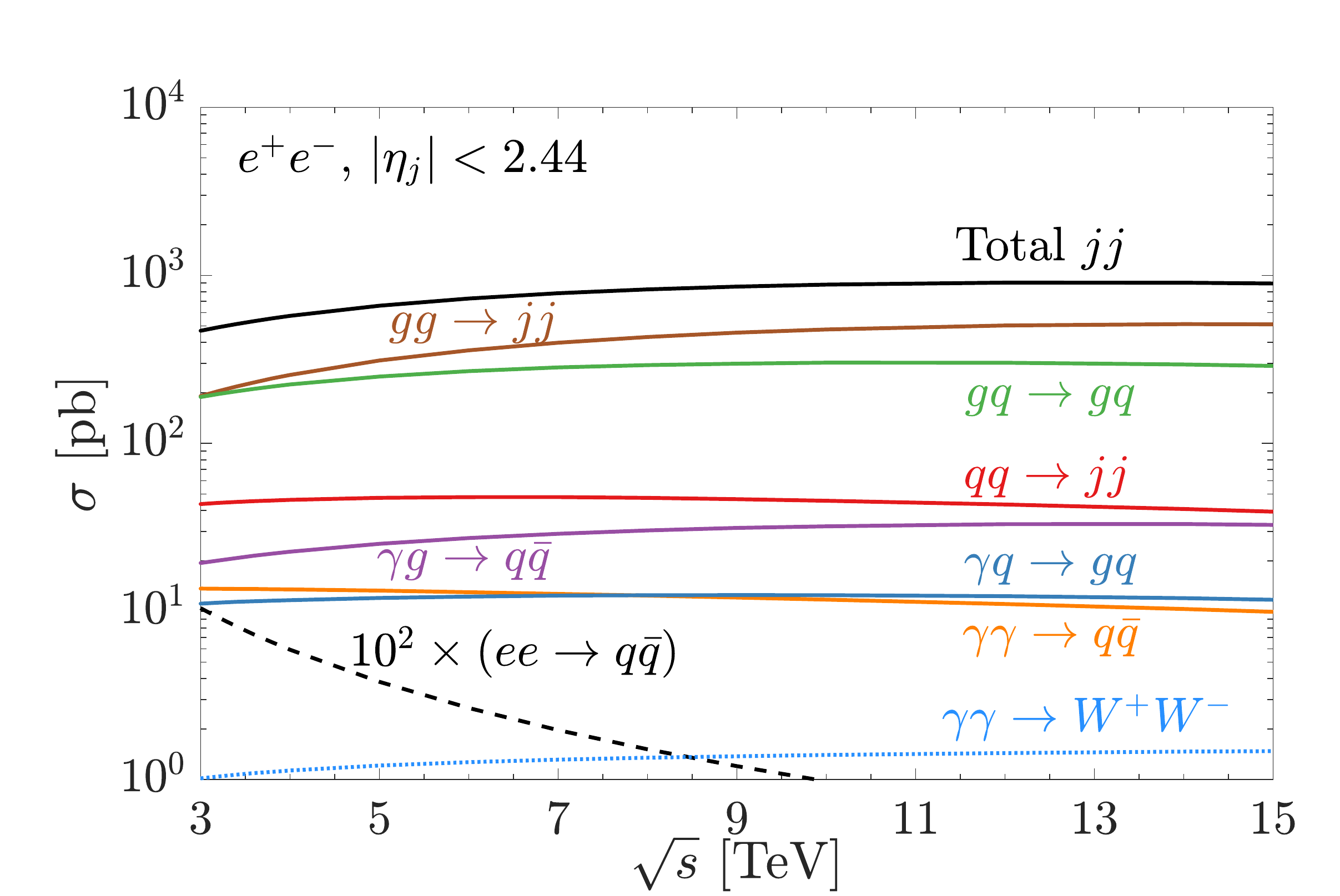}
	\includegraphics[width=.48\textwidth]{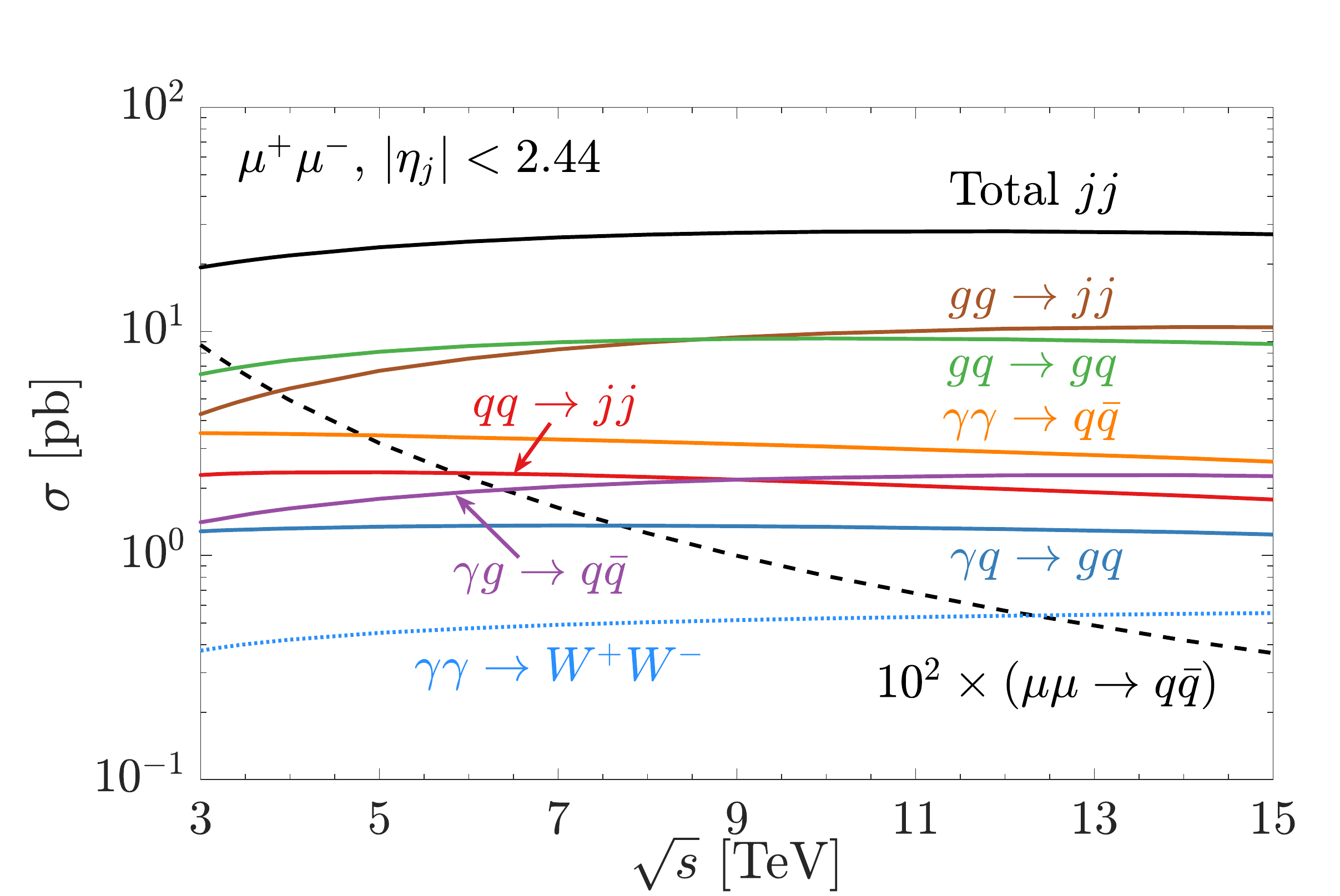}
	\caption{Cross sections for di-jet (or $W^+W^-$) production ($j=q,g$) versus the collider c.m.~energy for an $\ee$ collider (left panels) and a $\mm$ collider (right panels) with basic acceptance cuts in Eq.~(\ref{eq:cut}). 
	}
	\label{fig:dijet}
\end{figure}

Particularly important channels of our current interests are the jet production via the fusion mechanism, which would be the dominant phenomena at low $\sqrt{\hat s}$. The production channels include
\bea
&\gamma\gamma \to q\bar q,\ \gamma g \to q\bar{q},~\gamma q\to gq,\\
&q q\to qq\ (gg),\ gq\to gq\ {\rm and}\ gg\to gg\ (q\bar{q}),
\label{eq:jets}
\eea
where $q$ includes $d,u,s,c,b$ and the possible anti-quarks as well.
The PDFs and the corresponding partonic luminosities are already shown in Figs.~\ref{fig:PDFs} and \ref{fig:lumi} with the full DGLAP evolution at a double-log accuracy. 
We present the cross sections for di-jet production from initial states of photons, quarks, and gluons versus the collider c.m.~energy $\sqrt s = 3-15$ TeV at an $\ee$ collider (left panels) and a $\mm$ collider (right panels) in Fig.~\ref{fig:dijet}, subject to the acceptance cuts in Eq.~(\ref{eq:cut}) shown by the upper and lower panels. 
The patonic QCD jet cross sections are calculated at the leading order with \textsc{MadGraph5} v2.6.7 \cite{Alwall:2014hca} and cross-checked with \textsc{MCFM} v9.1 \cite{Campbell:2019dru} and \textsc{Sherpa} v2.2.10 \cite{Bothmann:2019yzt}. 

The standard factorization scale is chosen to be $Q=\sqrt{\hat s}/2$, while varying the scale to $Q=\sqrt{\hat{s}}$
gives a 6$\sim$15\% (30$\sim$40\%) enhancement of the cross sections for an $e^+e^-$ ($\mm$) collider, which characterizes the scale uncertainty.  
The rather large difference resulting from the scale choice is owing to the large $\alpha_s\log(Q^2)$ resummation.
It is important to note that, even originated from the photon splitting to quarks and then subsequently to gluons, the gluon and quark initiated processes exceed the photon fusion in the di-jet production rates by two (one) orders of magnitude for the electron (muon) collider. 
This is the result of large QCD resummation and the $g/q$ multiplicity. Depending on the acceptance cuts, the crossover of the $gg$ fusion to the $gq$ scattering happens around $3-4$ TeV for the electron collider and $8-12$ TeV for the muon collider. For the same reason, the $\gamma g \to jj $ process grows faster over the energy than the $\gamma\gamma \to jj$ fusion and takes over for the electron collider.
Compared with the photon-initiated processes, the angular dependence of the QCD jet cross sections is much stronger, due to the large QCD collinear logarithms $\alpha_s\log\theta^2$ effectively resummed by the DGLAP equations.

\begin{figure}[tb]
\centering
\includegraphics[width=.48\textwidth]{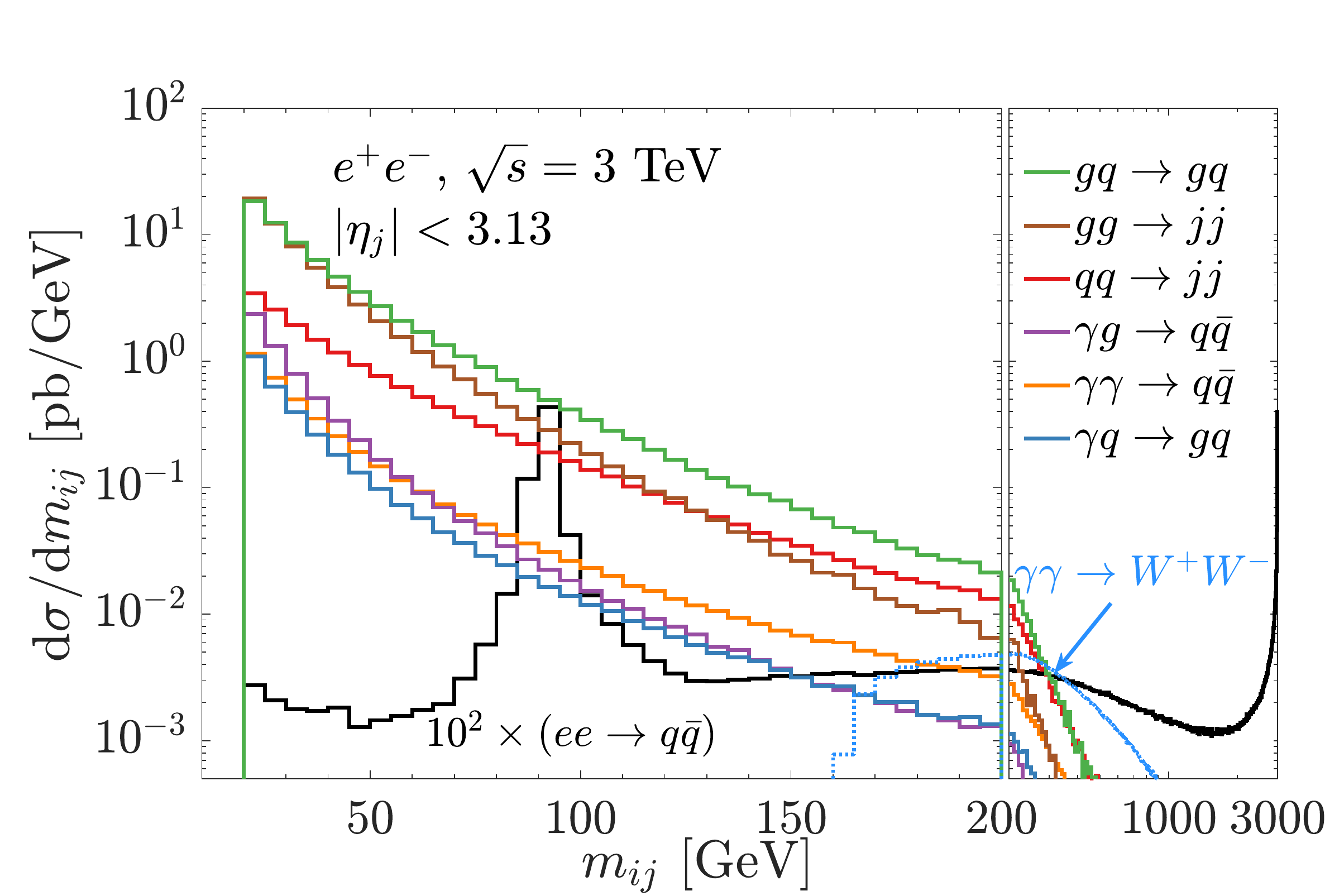}
\includegraphics[width=.48\textwidth]{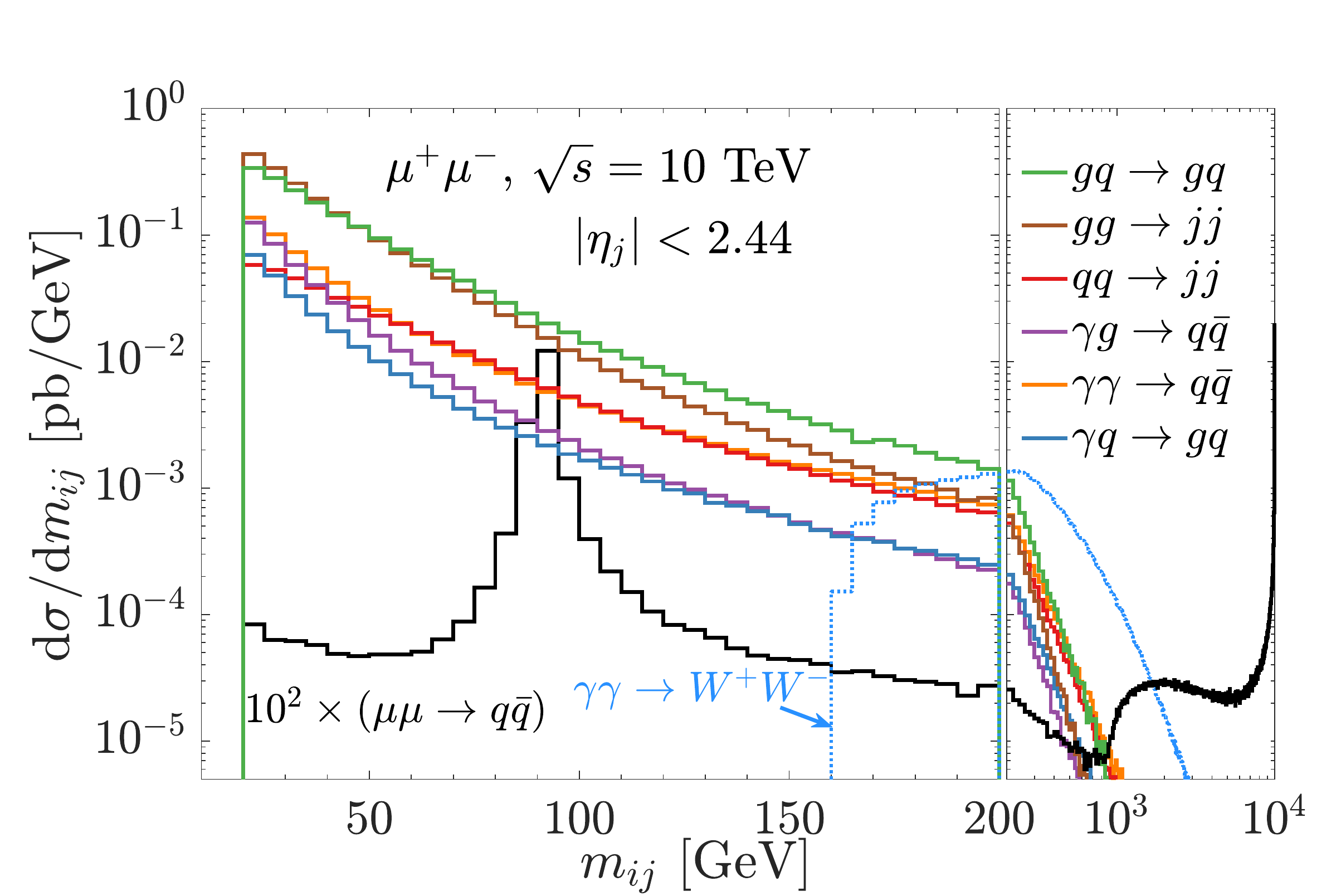}	
\includegraphics[width=.48\textwidth]{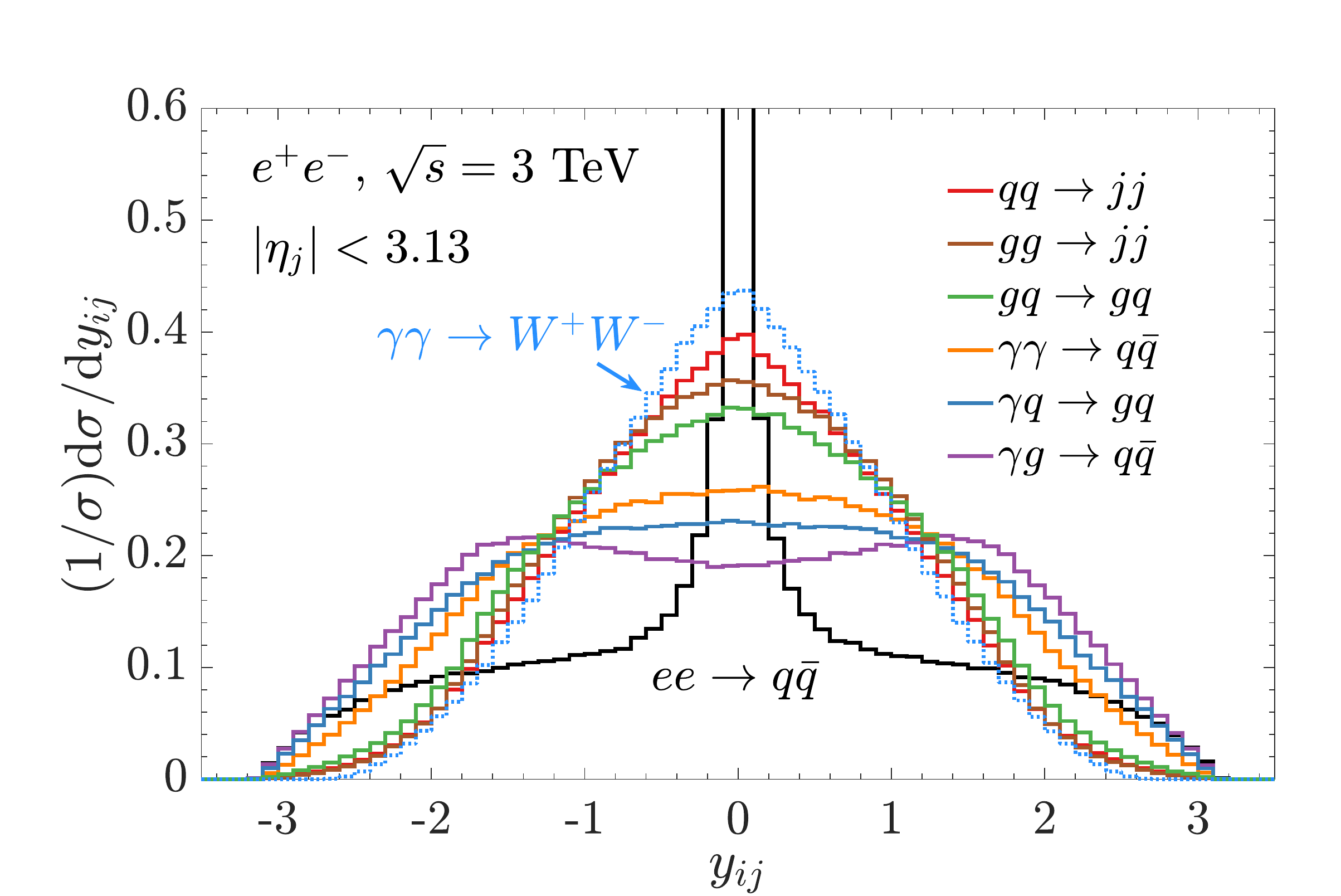}
\includegraphics[width=.48\textwidth]{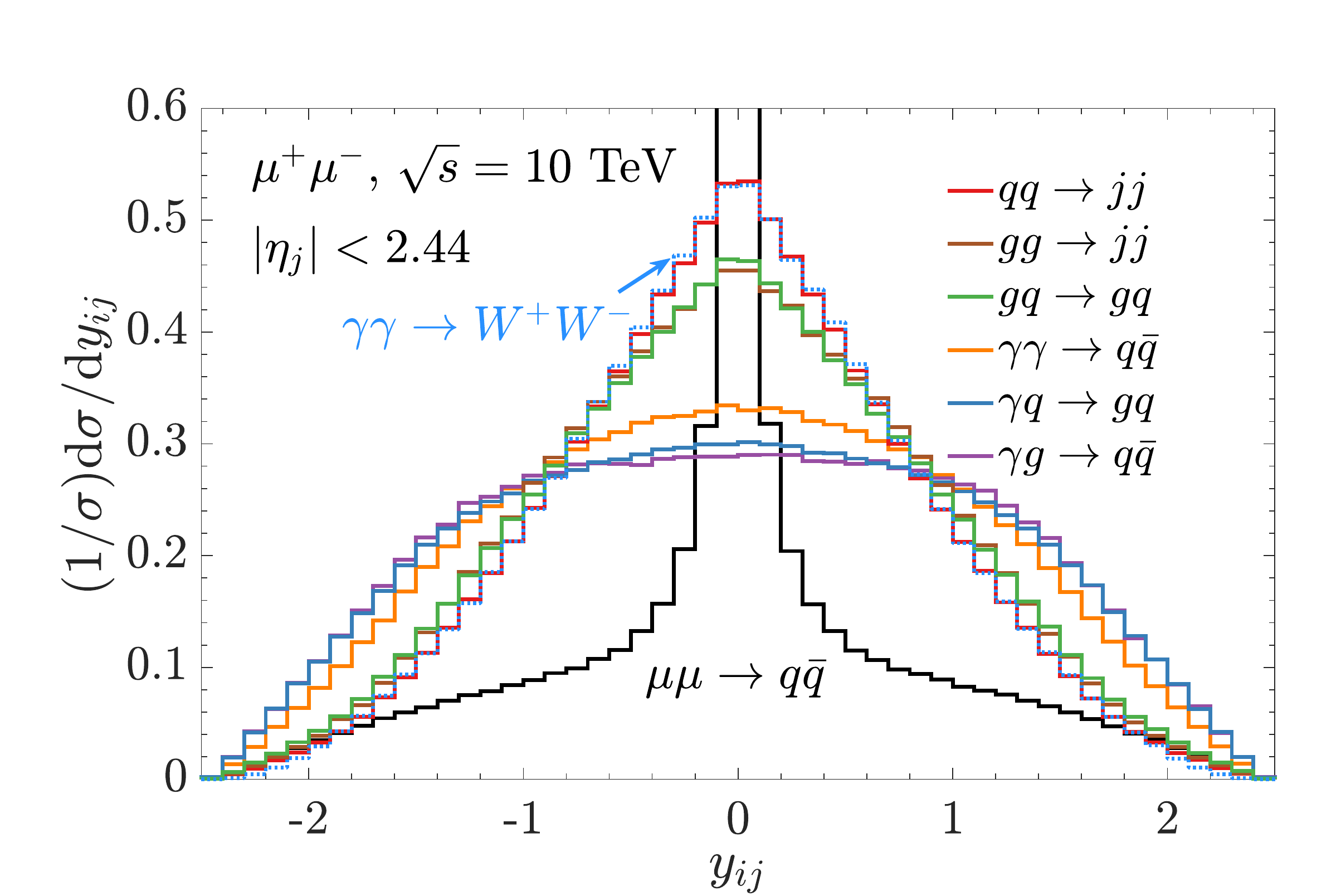}
\caption{Invariant mass ($m_{ij}$, upper panels) and rapidity ($y_{ij}$, lower panels) distributions for the di-jet (or $W^+W^-$) system from various sub-processes for an $e^+ e^-$ collider at $\sqrt s = 3$ TeV (left panels), and a $\mu^+\mu^-$ collider at $\sqrt s = 10$ TeV (right panels), respectively. }
\label{fig:mjjY}
\end{figure}

There are a number of improvements for the results shown here with respect to the QED calculations by EPA as in Fig.~\ref{fig:SMXS}. 
First, the higher-order cascade splittings $\gamma\to\ell^+\ell^-,q\bar{q}$ have been included, which will carry away a part of the momentum fraction from the initial photon and is roughly 5\% for an electron beam, and 3\% for a muon beam, estimated from Table \ref{tab:momavg}. 
Second, in our treatment of the full DGLAP evolution, the running effect of the QED coupling $\alpha(Q)$ is properly taken into account, with the boundary condition at the lepton mass set to be $\alpha(m_e^2)=1/137$ ($\alpha(m_\mu^2)=1/136$) and proper matching cross the mass thresholds. 
As expected, both effects tend to reduce the rate for photon-initiated processes with respect to the naive EPA calculations. 
As such, the cross section for $\gamma\gamma\to q\bar{q}$ receives about 16\% (8\%) reduction over the EPA results for electron (muon) colliders evaluated with the fixed value $\alpha=1/132.5$. 
Finally, we note that the other EW VBF contributions such as $\gamma Z,W^+W^-, W^\pm Z \to q\bar q'$ are sub-leading and contribute less than $1\%$, due to the suppression of the EW threshold above $M_Z$ or $2M_W$. 

One of the most striking aspects for a high-energy lepton collider is the combination of two characteristically different production mechanisms: the direct $e^+e^- / \mu^+\mu^-$ annihilation channels and the fusion processes. The former carries the full collider energy to reach a high threshold and the latter starts from the low energy to scan over the full spectrum. These distinctive kinematic features can be best shown by the invariant mass ($m_{ij}$) of the final state di-jet system as in the upper panels of Fig.~\ref{fig:mjjY} at $\sqrt s=3$ TeV for $\ee$ and 10 TeV for $\mm$, respectively. 
We see the clear separation of events from these two classes of reactions, 
peaked around the low threshold in $m_{ij}$ for the partonic fusion processes, and sharply peaked at the beam collision energy $\sqrt{s}$ for the annihilation process (a factor of 100 is multiplied here because of the smaller production rate). The long tail in low $m_{ij}$ for the annihilation process is due to the ISR, followed by another peek around the $Z$ resonance from the radiative return 
$\ell^+\ell^-\to Z\to jj$. 
In the 10 TeV $\mm$ collider case, the $m_{ij}$ distribution has a threshold kink around $m_{ij}\sim\sqrt{s}e^{-\eta}\approx870$ GeV, 
which is from the effect of the angular cut. This is not notable in the $\ee$ collider case with the cut $|\eta|<3.13$, because the location $m_{ij}\sim\sqrt{s}e^{-\eta}=130$ GeV is diluted by the falling from the resonant $Z$ peak. We also include a leading production channel $\gamma\gamma \to \ww$ in high-energy leptonic collisions for comparison. We see that the jet production is overwhelmingly larger until the kinematical region with a high invariant mass $m_{ij}\gtrsim200$ GeV.
The second distinctive kinematic feature manifests itself in the rapidity distributions of the di-jet system shown in the lower panels of Fig.~\ref{fig:mjjY} for $\ee$ and $\mm$, where the annihilation process is very central with back-to-back di-jets peaked at $y_{ij}\sim \log(x_1/x_2) \approx 0$, spreading out by the ISR. 
In comparison, the fusion process spread out, especially for the processes involving a photon due to the large imbalance between $x_1$ and $x_2$. The distribution for $\gamma\gamma \to \ww$ is also relatively more central. 

\begin{figure}[tb]
\centering	
\includegraphics[width=.48\textwidth]{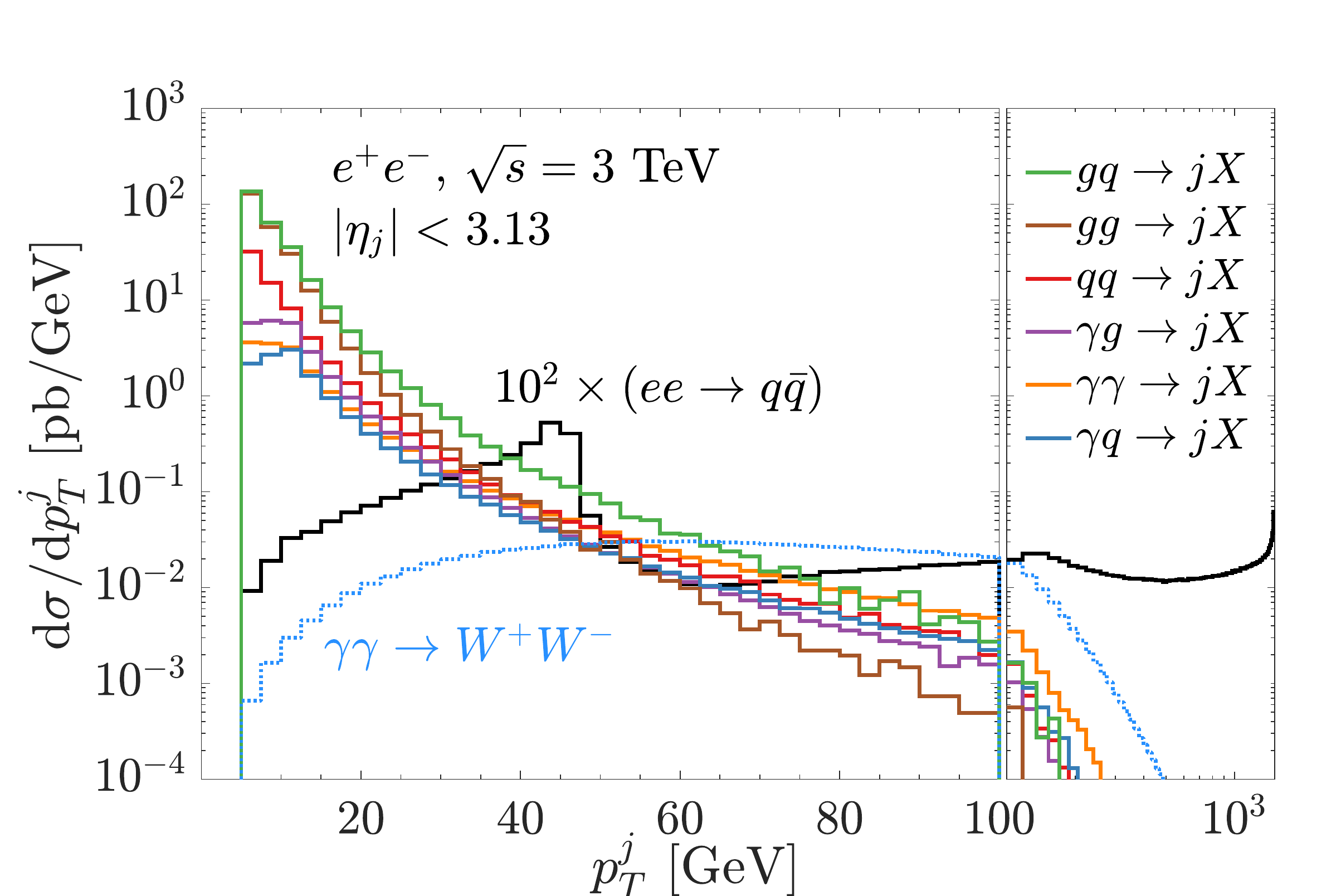}
\includegraphics[width=.48\textwidth]{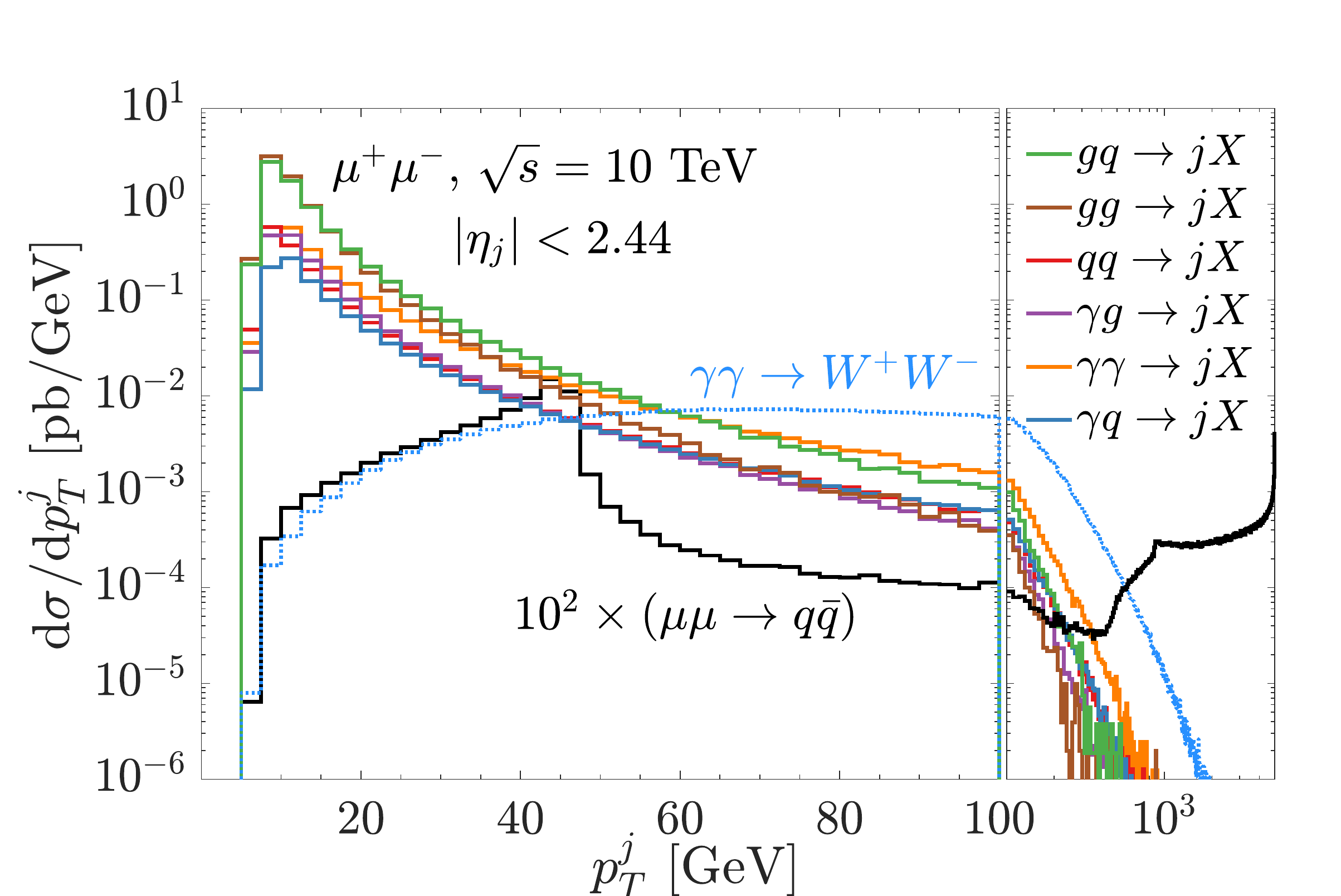}
\includegraphics[width=.48\textwidth]{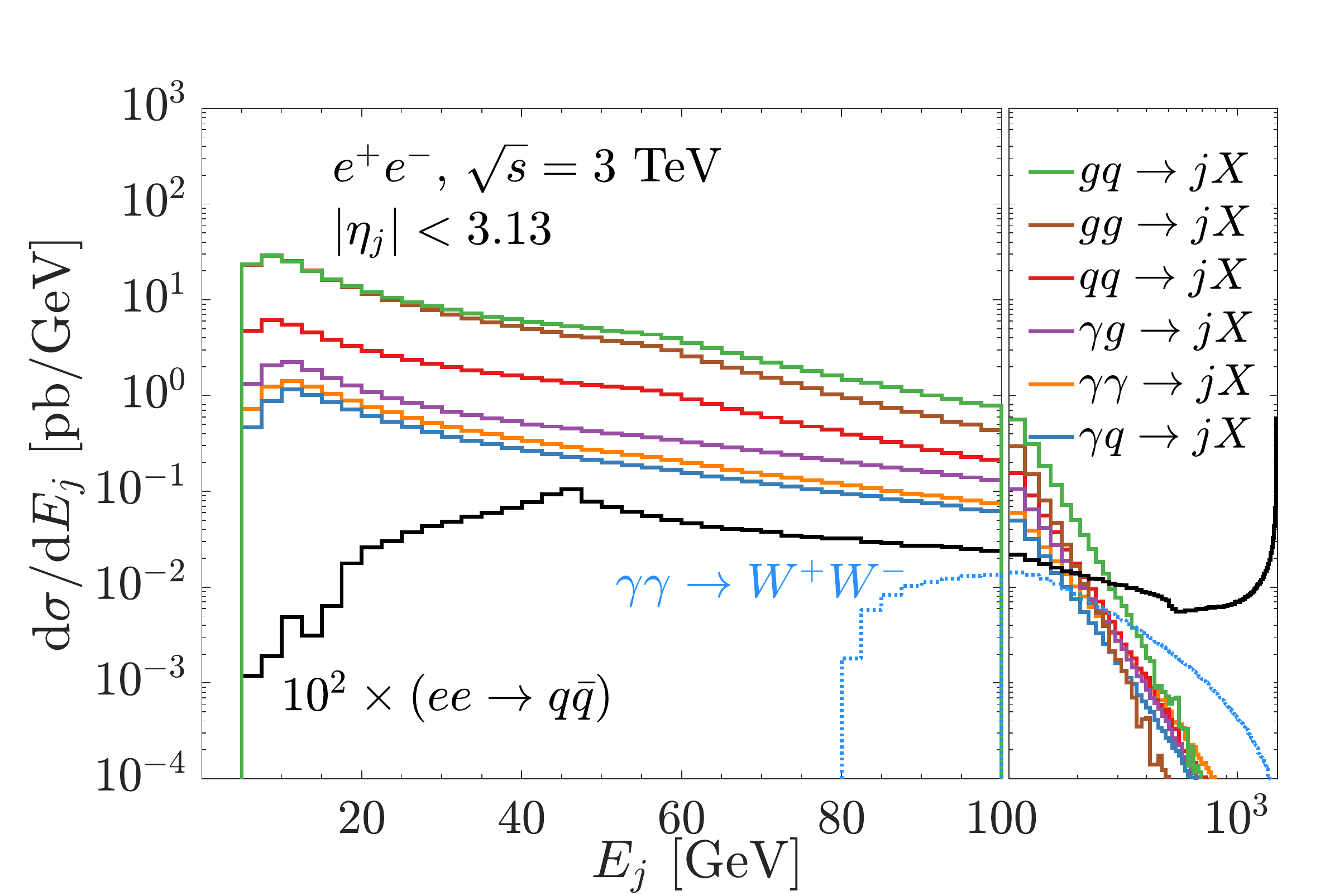}
\includegraphics[width=.48\textwidth]{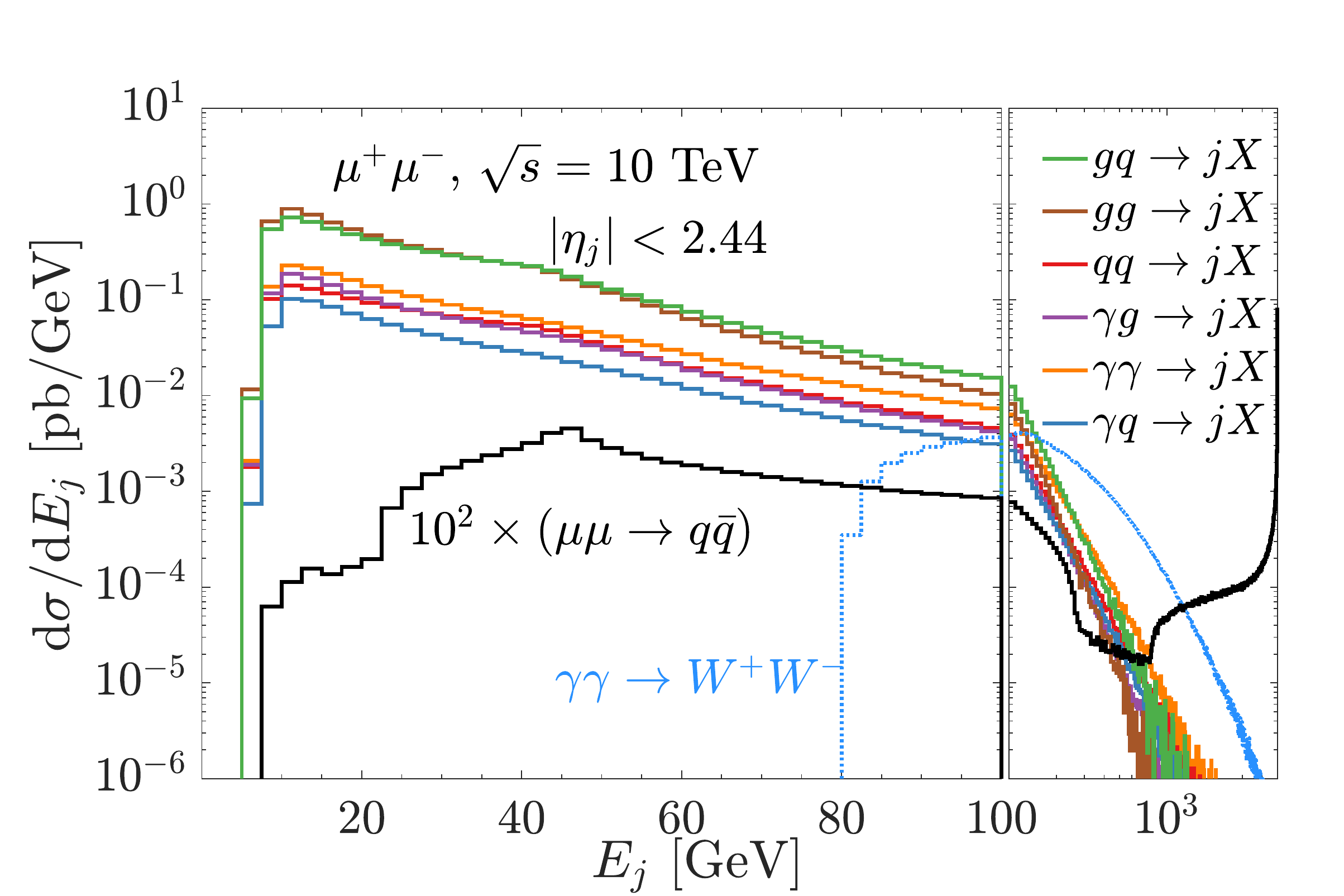}
\includegraphics[width=.48\textwidth]{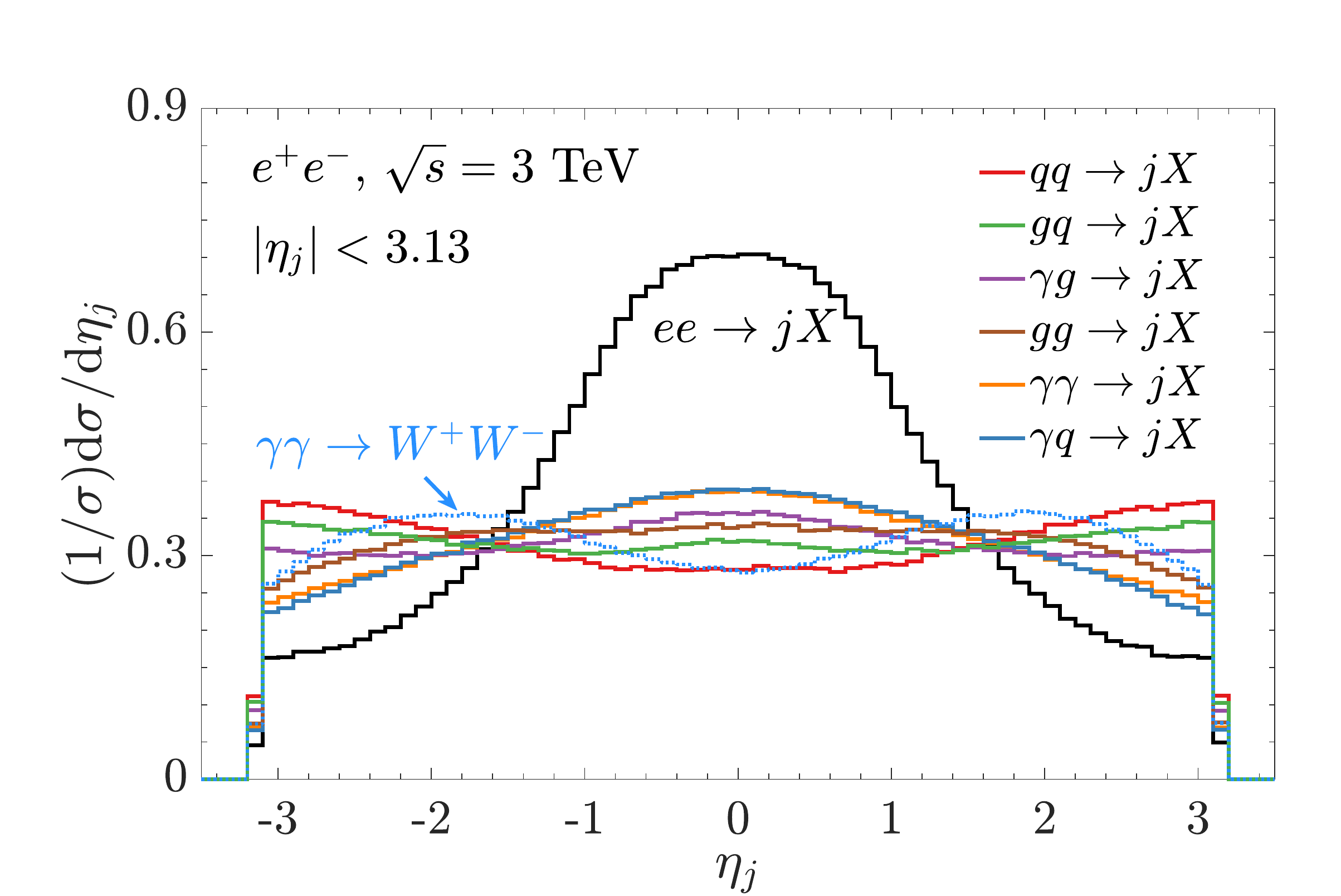}
\includegraphics[width=.48\textwidth]{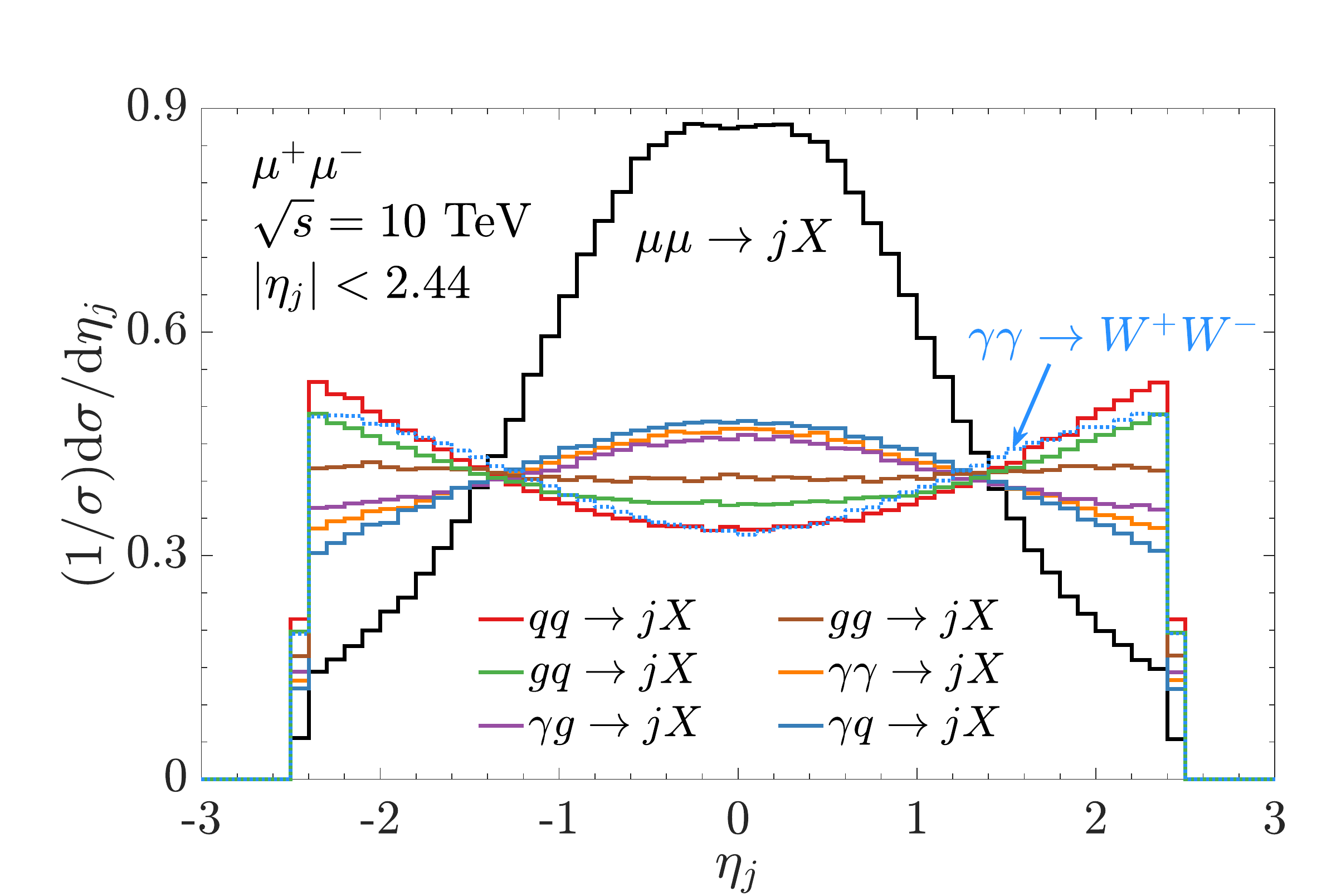}	
\caption{Inclusive jet (or $W$) distributions for transverse momentum ($p_T^j$, upper panels),
jet energy ($E_{j}$, middle panels) and the normalized 
pseudo-rapidity ($\eta_{j}$, lower panels) in various sub-processes for an $e^+ e^-$ collider at $\sqrt s = 3$ TeV (left panels), and a $\mu^+\mu^-$ collider at $\sqrt s = 10$ TeV (right panels), respectively.
}
\label{fig:distribution}
\end{figure}

Finally, we present some kinematic distributions of the inclusive jets in Fig.~\ref{fig:distribution}, 
the transverse momentum ($p_T^j$, upper panels), the jet energy ($E_j$, middle panels), and the pseudo-rapidity ($\eta_j$, lower panels), at a 3 TeV $\ee$ (left panels) and a 10 TeV $\mm$ (right panels) collider, respectively.\footnote{We remind the reader that inclusive jets include any jets in an event. That is to say, each di-jet event is counted twice.}
The $p_T^j$ distributions in Fig.~\ref{fig:distribution} resemble very similar features as those of $m_{ij}$ in Fig.~\ref{fig:mjjY}, with the Jacobian peaks around the $p_T^j\sim m_{ij}/2$ for the fusion processes, and peaked sharply at $\sqrt s /2$ and $M_Z/2$ for the annihilation processes. 
We once again see the dominant QCD jet production over the $\ww$ channel until 
the kinematical region with a high transverse momentum $p_T^j >60$ GeV. 
We note that there is a peculiar peak structure in the $p_T^j$ distribution for the annihilation processes. After the peak at $\sqrt{s}/2$, it falls and rises again around $p_T^j\sim\sqrt{s}e^{-\eta}=130~(870)$ GeV, the same location as the $m_{ij}$ kink.
The dip around 300 GeV for the 10 TeV $\mm$ collider case is just the cross point between the falling from the Jacobi peak $M_Z/2$ and rising to the cut point $\sqrt{s}e^{-\eta}$.
Furthermore, we see from the energy distributions that the $\ww$ channel takes over after its energy above 400 GeV (200 GeV) for the $\ee$ collider ($\mm$ collider). 
The inclusive pseudo-rapidity distributions in Fig.~\ref{fig:distribution} demonstrate that the QCD partonic contributions are mostly forward-backward, while those of $\gamma\gamma$ and $\gamma q(g)$ are more isotropic, and the 2-body annihilation process via an $s$-channel vector boson exchange presents the typical $(1+\cos^2\theta)$ distribution.

\section{Summary and conclusions}
\label{sec:sum}

In high-energy leptonic collisions, such as at a multi-TeV muon collider, the collinear splittings of electroweak gauge bosons and leptons are the dominant phenomena, and thus the scattering processes should be formulated in terms of the EW parton distribution functions (EW PDFs). 
We complete this formalism in the Standard Model to include the QCD sector and evaluate the quark and gluon PDFs inside a lepton by solving the fully-coupled DGLAP equations at the double-log accuracy, as presented in Sec.~\ref{sec:PDF}. We see that, dominantly from the photon splitting, there are significant gluon and quark contents in high energy lepton beams as shown in Figures \ref{fig:PDFs} and \ref{fig:lumi}. In comparison, while the photon PDF in an electron is larger than that in a muon by about a factor of two below the EW scale, the quark/gluon PDFs are substantially larger in an electron than that in a muon due to the large log resummation from QCD splittings. The subsequent splittings also make a notable effect as ISR on the lepton beam profile. 
The initial state of quarks and gluons will lead to QCD processes with large cross sections and will dominate the overall event shape in high-energy leptonic collisions with low and moderate transverse momenta. They may also induce the production of new colored particles \cite{Han:2010rf}. 

\begin{table}
\centering
	\begin{tabular}{|c|c|c|c|c|c|}
		\hline
		$\ee$ [$\sqrt{s}$] & $\sigma$ [pb]
		$jj$ & $\ee$ & $\tau^+\tau^-$& $\ww$ &   $t\bar{t}$\\
		\hline
		3 TeV & 800(470) &33(15) & 16(11) & 1.9(1.2) & 0.035(0.032) \\
		6 TeV & 1200(730) &19(11) & 15(10) & 2.3(1.3) & 0.023(0.019) \\
		10 TeV& 1400(880) &15(9.5)& 13(9.1) & 2.5(1.4) & 0.023(0.017) \\
		14 TeV& 1400(910) &12(8.2)& 11(8.0) & 2.7(1.5) & 0.024(0.017) \\
		\hline
	\end{tabular}
	\begin{tabular}{|c|c|c|c|c|c|}
		\hline
		$\mm$ [$\sqrt{s}$] & $\sigma$ [pb]
		$jj$ & $\mm$ & $\tau^+\tau^-$ &$\ww$  &  $t\bar t$ \\
		\hline
		3 TeV & 34(19) & 21(6.9) & 4.1(2.7) & 0.82(0.52) & 0.027(0.025) \\
		6 TeV & 43(25) & 8.3(3.7)& 3.9(2.6) & 0.89(0.51) & 0.012(0.011) \\
		10 TeV & 46(28) & 5.1(2.7)& 3.5(2.4) & 0.97(0.54) & 0.010(0.0078) \\
		14 TeV & 45(28) & 3.8(2.3)& 3.0(2.1) & 1.0(0.56)  & 0.010(0.0073) \\
		\hline
	\end{tabular}
	\caption{Some representative cross sections in $\ee$ and $\mm$ collisions including both annihilation and fusion for a variety of energies. We have included the ISR for the annihilation processes. The fusion to $W^+W^-,t\bar{t}$ cross sections only include the dominated $\gamma\gamma$ initialized processes with the resummed $\gamma$ PDF. The acceptance cuts in Eq.~(\ref{eq:cut}) are applied to the final-state particles, including the $W^+W^-$ and $t\bar{t}$ as well. The numbers outside (inside) of the parentheses correspond cross sections with the acceptance cut $|\eta_j|<3.13$ ($|\eta_j|<2.44$).} 
	\label{tab:xsec}
\end{table}

In Sec.~\ref{sec:SMProcesses}, we studied the production cross sections in our PDF framework. We compared the standard QED processes in leptonic collisions at multi-TeV energies and showed the dominance of the fusion mechanism in Fig.~\ref{fig:SMXS}. We then gave the prediction for jet production of quarks and gluons in Fig.~\ref{fig:dijet}. We found that, as expected, the QCD jet production initiated by $q/g$ yields the dominant processes, about two orders (one order) of magnitude larger than the EW fermion pair production at an $\ee$ ($\mm$) collider, reaching a large production rate of about 1 nb (50 pb), with a moderate acceptance cut. 
We summarize some representative cross sections in $\ee$ ($\mm$) collisions for a variety of energies in Table \ref{tab:xsec}.
The total cross sections include both annihilation and fusion processes. The fusions to $W^+W^-$ and $t\bar{t}$ only include the dominated $\gamma\gamma$ initialized processes \cite{Han:2020uid}.
The kinematic cuts in Eq.~(\ref{eq:cut}) are employed to the $W$ boson and top quarks, as well. 
To have a more complete picture with respect to the hadronic production at low scattering energies, 
we also calculated the total cross section for the photon-induced hadronic production adopting two models: Pythia and ``SLAC'', as shown in Fig.~\ref{fig:hadron}. We see that the cross sections can reach the level of one hundred  (a few tens) of nano-barns at high-energy electron (muon) colliders. Although the rate for the hadronic production 
is high, the events populate in the low $p_T$ region typically below a few GeV. 

Of particular interests are the differential distributions for di-jet system in Fig.~\ref{fig:mjjY}, and for jet-inclusive in Fig.~\ref{fig:distribution}. The general features emerge again that the $\ee\ (\mm)$ annihilation is mostly central with $\hat{s}\approx s$, the fusion processes populate at $\sqrt{\hat{s}}\approx m_{ij}$, and QCD jet production dominates up to $p_T^j \approx 60$ GeV.
Since the events tend to populate near the threshold, the photon splitting governs the fate, especially below the EW scale, while the heavy EW gauge bosons do lead to substantial contribution at high scales.

As a final remark, our approach to the quark/gluon PDFs induced by the EW interactions is equally applicable to hadronic collisions with quarks as the radiation source. Since the simulations for photon-induced high-$p_T$ jet events from perturbative QCD calculations do not exist in the current event-generator packages, our formalism should be adopted by the event generators to simulate SM processes and the leading QCD backgrounds at lepton colliders.

\begin{acknowledgments}
We thank Manuel Drees, Rohini Godbole, and Daniel Schulte for discussions on the $\gamma\gamma$-induced hadronic production. 
This work was supported in part by the U.S.~Department of Energy under grant No.~DE-SC0007914, U.S. National Science Foundation under Grant No.~PHY-1820760, and in part by the PITT PACC.
\end{acknowledgments}

\appendix

\section{Solving the DGLAP equations: the valence non-singlet PDF as an example}
\label{app:solution}
In this appendix, we take the non-singlet PDF of valence lepton as an example to show the approach we developed to solve the DGLAP equations numerically. The more comprehensive details related to the singlet, photon, gluon PDFs, and high-energy EW ones above $\muEW$ are beyond the content of this paper, which are left for a future publication \cite{ours}. We will start with an approximate analytical solution and address its drawbacks. Afterward, we will move to the technicalities to tackle them numerically.

\subsection{An approximate analytical solution}
\label{app:largeX}
As discussed in the Sec. \ref{sec:QCD}, the DGLAP equation for the non-singlet PDFs are given as Eq. (\ref{eq:DGLAPNS}). The only non-trivial initial condition is for the valence flavor PDF, as Eq. (\ref{eq:NSval}). The corresponding splitting function is 
\begin{equation}
P_{\ell\ell}(x,Q^2)=a\left[\frac{1+x^2}{(1-x)_+}+\frac{3}{2}\delta(1-x)\right],
\end{equation}
where $a=\alpha/(2\pi)$ and the plus (+) prescription is defined as
\begin{equation}
\int_{x}^{1} \mathrm{~d} z f(z)[g(z)]_{+}=\int_{x}^{1} \mathrm{~d} z f(z) g(z)-f(1) \int_{0}^{1} \mathrm{~d} z g(z).
\end{equation}
With the Mellin transform,
\begin{equation}\label{eq:Mellin}
\tilde{f}(N)=\mathcal{M}[f]=\int_0^1\dd x x^{N-1}f(x),
\end{equation}
the convolution of Eq. (\ref{eq:conv}) becomes a multiplication
\begin{equation}
\mathcal{M}[f\otimes g]=\tilde{f}(N)\tilde{g}(N),
\end{equation}
where the symbol with a tilde specifies the same variable in the Mellin-$N$ space.
The DGLAP equation for the non-singlet in Eq. (\ref{eq:NSval}) turns into 
\begin{equation}\label{eq:NSMN}
\dv{\tilde{f}_{\ell}(N,Q^2)}{L}=\tilde{P}_{\ell\ell}(N,Q^2) \tilde{f}_{\ell}(N,Q^2),
\end{equation}
where $L=\log Q^2$. For simplification, we leave out the ``NS" label and only refer to the valence flavor in this section. The specific expression for initial condition and splitting function in the Mellin-$N$ space are
\begin{equation}
\tilde{f}_{\ell}(N,m^2_{\ell})=1, ~ \tilde{P}_{\ell\ell}(N,Q^2)=a\left[\frac{3}{2}-S_1(N-1)-S_1(N+1)\right],
\end{equation}
where $\gamma_E$ is the Euler constant. The $S_{m}$ is the harmonic series defined as
\begin{equation}
S_{m}(N)=\sum_{i=1}^{N}\frac{1}{i^{m}}.
\end{equation}
The $S_1(N)$ can be analytically continued as
\begin{equation}
S_1(N)=\gamma_E+\psi(N+1), ~ \psi(z)=\dv{z}\ln\Gamma(z).
\end{equation}
With neglecting the running of the coupling, \emph{i.e.}, a fixed $a$ value, the splitting function is independent on $Q^2$, \emph{i.e.}, $\tilde{P}_{\ell\ell}(N,Q^2)=\tilde{P}_{\ell\ell}(N)$. Eq. (\ref{eq:NSMN}) can be solved analytically as
\begin{equation}\label{eq:solu1}
\tilde{f}_{\ell}(N,Q^2)=\exp{\tilde{P}_{\ell\ell}(N)L}.
\end{equation}

In the large-$N$ limit ($N\to\infty$), which corresponds to a large $x$ ($x\to1$), we have an approximation $S_1(N)\approx\gamma_E+\log N$. Therefore, the splitting function becomes 
\begin{equation}
\tilde{P}_{\ell\ell}(N)\approx a\left[\frac{3}{2}-2(\gamma_E+\log N)\right]
\end{equation}
The solution in Eq. (\ref{eq:solu1}) can be simplified as
\begin{equation}\label{eq:NSsolu}
\tilde{f}_{\ell}(N,Q^2)\approx\exp{a L\left(\frac{3}{2}-2\gamma_E\right)-2aL\log N}=e^{\beta\lambda}N^{-\beta},
\end{equation}
where
\begin{equation}
\beta=2 aL, ~\lambda=\frac{3}{4}-\gamma_E.
\end{equation}
With the known Mellin transform
\begin{equation}
\mathcal{M}\left[(1-x)^{-1+\kappa}\right]=\frac{\Gamma(\kappa)\Gamma(N)}{\Gamma(\kappa+N)}\xrightarrow{N\to\infty}\Gamma(\kappa)N^{-\kappa},
\end{equation}
we can convert the large-$N$ solution analytically back to the $x$ space as
\begin{equation}\label{eq:large-x}
f_{\ell}(x, Q^2)\approx\frac{e^{\beta\lambda}}{\Gamma(\beta)}(1-x)^{-1+\beta}.
\end{equation}
Up to this stage, we have obtained the all-order resummation of the non-singlet PDF of valence lepton in the large-$x$ limit, as in Refs. \cite{Gribov:1972ri,Bertone:2019hks}.

Nevertheless, we want to remind the reader that two assumptions are critical to this solution: the large $N$ limit and a fixed coupling $a$. The large-$N$ limit means that the solution Eq. (\ref{eq:large-x}) is only reliable in the large-$x$ limit, which will be violated at a non-trivial $x$ value ($x<1$). In addition, the fixed coupling assumption also restricts its applicability. In QED, it is not a big problem, as the fine-structure coupling $\alpha$ runs slowly with energy. However, this is not the case anymore in QCD, where the strong coupling $\alpha_s$ is much larger and runs drastically, especially in the low-energy region around $Q\gtrsim\muQCD$. Moreover, obtaining a simple solution form as Eq. (\ref{eq:solu1}) is not possible for the coupled equation, Eq. (\ref{eq:SG}), of the singlet, photon, and gluon PDFs. These three reasons drive us to an alternative numerical approach, which will be described in the next subsection. 

\subsection{A numerical solution}
In this subsection, we will still constrain us within the non-singlet valence PDF. However, the techniques we developed are equally applicable to the singlet, photon, and gluon as well. Just because of the complexity of the coupled matrix equation, we leave the details to a future work \cite{ours}.

{\bf Running couplings.}
As we mentioned above, the couplings $\alpha, ~\alpha_s$ run as well with the energy, similarly to the $\alpha_1,\alpha_2$ above $\muEW$. We take the leading order running couplings as 
\begin{equation}\label{eq:runAlfa}
\alpha(Q^2)=\frac{\alpha(m_{\ell}^2)}{1+\beta_e\frac{\alpha(m_\ell^2)}{2\pi}\log(Q^2/m_\ell^2)}, ~
\alpha_s(Q^2)=\frac{4\pi}{\beta_s\log(Q^2/\Lambda_{\rm QCD}^2)},
\end{equation}
where 
\begin{equation}
\beta_e=-\frac{2}{3}(N_{\ell}+N_cN_u e_u^2+N_cN_de_d^2), ~ \beta_s=11-\frac{2}{3}(N_u+N_d).
\end{equation}
which match to the leading splitting functions. The corresponding numerical inputs are taken as
\begin{equation}
\alpha(m_e^2)=\frac{1}{137}~\textrm{or}~\alpha(m_\mu^2)=\frac{1}{136}, ~\Lambda_{\rm QCD}=89.9~\MeV,
\end{equation}
which ensure $\alpha(M_Z^2)=1/128.8$ and $\alpha_s(M_Z^2)=0.118$. We have performed the complete matchings whenever crossing a heavy-flavor threshold in the $\alpha_e$ running, while $\alpha_s$ in the low-$Q^2$ region is taken as an extrapolation.

{\bf Iterations.}
In the Mellin-$N$ space, the non-singlet valence PDF satisfies the evolution in Eq. (\ref{eq:NSMN}), with $a$ running with scale as well. 
This equation can be numerically solved in terms of the Euler method. 
Suppose we want to obtain the $\tilde{f}_{\ell}$ at a given scale $Q$ with an initial condition at $Q_0=m_\ell$. We can divide the running parameter $L=\log(Q^2/m_{\ell}^2)$ into $K$ steps, with step length $h=L/K$ and the $L_k$ grid as 
\begin{equation}
L_{k}=kh, ~k=0,\cdots, K.
\end{equation}
Other quantities at the $k^{\rm th}$ grid can be determined correspondingly as
\begin{equation}
Q_k^2=m_{\ell}^2\exp(L_k), ~a_k=a(Q_k^2), ~\tilde{P}_k=\tilde{P}_{\ell\ell}(N,Q_k^2).
\end{equation}
Then, the differential equation, Eq. (\ref{eq:NSMN}), can be approximated as
\begin{equation}
\frac{\Delta \tilde{f}_{\ell}}{\Delta L}
=\frac{\tilde{f}_{k}-\tilde{f}_{k-1}}{h}
=\tilde{P}_{k-1}\tilde{f}_{k-1}+\mathcal{O}(h), 
~k=1,\cdots, K,
\end{equation}
where $\tilde{f}_k=\tilde{f}_{\ell}(N,Q_k^2)$ and $\mathcal{O}(h)$ is the local truncation error (LTE).
In such a way, we get an iteration equation,
\begin{equation}\label{eq:Euler}
\tilde{f}_{k}=\tilde{f}_{k-1}+\tilde{P}_{k-1}\tilde{f}_{k-1}h+\mathcal{O}(h^2).
\end{equation}
Therefore, the final numerical solution can be obtained through $K$ steps of iterations as 
\begin{equation}\label{eq:soluK}
\tilde{f}_{\ell}(N,Q^2)=\tilde{f}_K=\tilde{f}_0\prod_{k=1}^{k=K}(1+\tilde{P}_{k-1} h)+\mathcal{O}(h).
\end{equation}
When the step number $K$ is large enough, the global truncation error (GTE), $\mathcal{O}(h)$, can be negligible, and the solution $\tilde{f}_{K}$ will converge to its true value $\tilde{f}_{\ell}(N,Q^2)$. If we ignore the running of the coupling, the splitting function is a constant, $\tilde{P}_k=\tilde{P}_{\ell\ell}(N)$, which does not depend on the grid $Q_k^2$. The solution in Eq (\ref{eq:soluK}) becomes 
\begin{equation}
\tilde{f}_K=\left[1+\tilde{P}_{\ell\ell}(N)h\right]^{K}=\left[1+\tilde{P}_{\ell\ell}(N)\frac{L}{K}\right]^{K}
\xrightarrow{K\to\infty}\exp{\tilde{P}_{\ell\ell}(N)L},
\end{equation}
in which we have substituted the initial condition $\tilde{f}_0=\tilde{f}_\ell(N,m_\ell^2)=1$ already. We see the iteration in the large $K$ limit reproduces the analytical solution in Eq. (\ref{eq:solu1}).

The Euler method in Eq. (\ref{eq:Euler}) corresponds to the 1$^{\rm st}$ order of Runge-Kutta (RK) algorithm. The convergence can be improved with higher-order corrections. In practice, we employ the 4$^{\rm th}$ order of RK approach in our real implementation:
\begin{equation}
\begin{array}{l}
\tilde{f}_{k}=\tilde{f}_{k-1}+\frac{1}{6} h\left(k_{1}+2 k_{2}+2 k_{3}+k_{4}\right)+\mathcal{O}\left(h^{5}\right),
\end{array}
\end{equation}
with 
\begin{equation}
\begin{aligned}
&k_{1}=\tilde{P}_{k-1} \tilde{f}_{k-1}, ~
&&k_{2}= \tilde{P}_{k-1 / 2}\left(\tilde{f}_{k-1}+h k_{1} / 2\right) \\
&k_{3}=\tilde{P}_{k-1 /2}\left(\tilde{f}_{k-1}+h k_{2} / 2\right), ~
&&k_{4}=\tilde{P}_k\left(\tilde{f}_{k-1}+h k_{3}\right)
\end{aligned}
\end{equation}
where $\tilde{P}_{k-1/2}$ is evaluated at $L_{k-1/2}=(k-1/2)h$. The corresponding GTE is $\mathcal{O}(h^4)$.

We notice that in Ref. \cite{Bertone:2019hks},  Frixione \emph{et al.} took a perturbative expansion of the QED PDFs, which gives an equivalent solution. In this framework, we expand $\tilde{f}_{\ell}$ in term of a series of $L$ as
\begin{equation}
\tilde{f}_{\ell}=\sum_{k=0}^{\infty}\tilde{f}_{\ell}^{(k)}L^k.
\end{equation}
Similarly with a fixed coupling assumption, Eq. (\ref{eq:NSMN}) becomes 
\begin{equation}
\sum_{k=0}^{\infty}\tilde{f}_{\ell}^{(k)}kL^{k-1}=\tilde{P}_{\ell\ell}(N)\sum_{k=0}^{\infty}\tilde{f}_{\ell}^{(k)}L^k.
\end{equation}
By matching the coefficient of $L^k$ order by order, we get the a recursive relation of coefficients,
\begin{equation}
\tilde{f}_{\ell}^{(k)}=\frac{1}{k}\tilde{P}_{\ell\ell}(N)\tilde{f}_{\ell}^{(k-1)}.
\end{equation}
Therefore, we can obtain a expansion solution as
\begin{equation}
\tilde{f}_{\ell}
=\sum_{k=0}^{\infty}\frac{1}{k!}\left[\tilde{P}_{\ell\ell}(N)L\right]^k
=\exp{\tilde{P}_{\ell\ell}(N)L},
\end{equation}
which also returns to Eq. (\ref{eq:solu1}). 

Similar to our iteration approach, the perturbative expansion can also deal with the running coupling $\alpha$, already demonstrated in Ref. \cite{Bertone:2019hks}. We also realize that this expansion converges faster than our iteration in the QED evolution. However, this approach only works efficiently when the DGLAP equation only involves one coupling, so that we can effectively replace the running parameter $L$ or $Q^2$ with the running coupling or an equivalent variable. In the case of the QED and QCD mixed DGLAP equation, the expansion method loses its efficiency, mostly because the QED and QCD couplings run differently. The QED coupling $\alpha$ increases with scale, while the QCD one $\alpha_s$ decreases. If we naively take $L$ as the expansion parameter, the strong coupling $\alpha_s$ oscillates with the expansion order increasing, which will significantly hamper the efficiency of convergence. For this reason, we will stick with our iteration approach, which applies to the QCD-involved singlet, photon, and gluon PDFs, as well as EW PDFs above $\muEW$ \cite{ours}.

{\bf Mellin Inversion -- Talbot algorithm.} 
Up to now, we have obtained the solution of DGLAP solution in the Mellin $N$ space, based on the iteration approach, as outlined above.  The next target is to invert the Mellin-$N$ solution back to the $x$ space. In App. \ref{app:largeX}, we demonstrated an example to invert the large-$N$ solution analytically. However, this method is only applicable in a few limited cases when we know the analytical Mellin inversion form. In general, the analytical Mellin inversion is not possible, and we have to rely on a numerical evaluation. The inversion of the Mellin transform of Eq. (\ref{eq:Mellin}) reads
\begin{equation}\label{eq:Inversion}
f(x)=\mathcal{M}^{-1}[\tilde{f}]=\frac{1}{2 \pi i} \int_{c-i \infty}^{c+i \infty} \mathrm{d} N x^{-N} \tilde{f}(N),
\end{equation}
where the real number $c$ is arbitrary as long as it meets certain conditions.\footnote{In terms of the Mellin inversion theorem, these conditions include: (1) $\tilde{f}(N)$ is analytic in the strip $\Re N\in(a,b)$; (2) $\tilde{f}(N)|_{\Im N\to\pm\infty}\to0$ uniformly for any real value $c\in(a,b)$; (3) the integral in Eq.~(\ref{eq:Inversion}) is converging absolutely.} 
This integration can be performed in terms of the Talbot algorithm \cite{Abate:2004a}. 
\begin{figure}
	\center
	\includegraphics[width=0.45\textwidth]{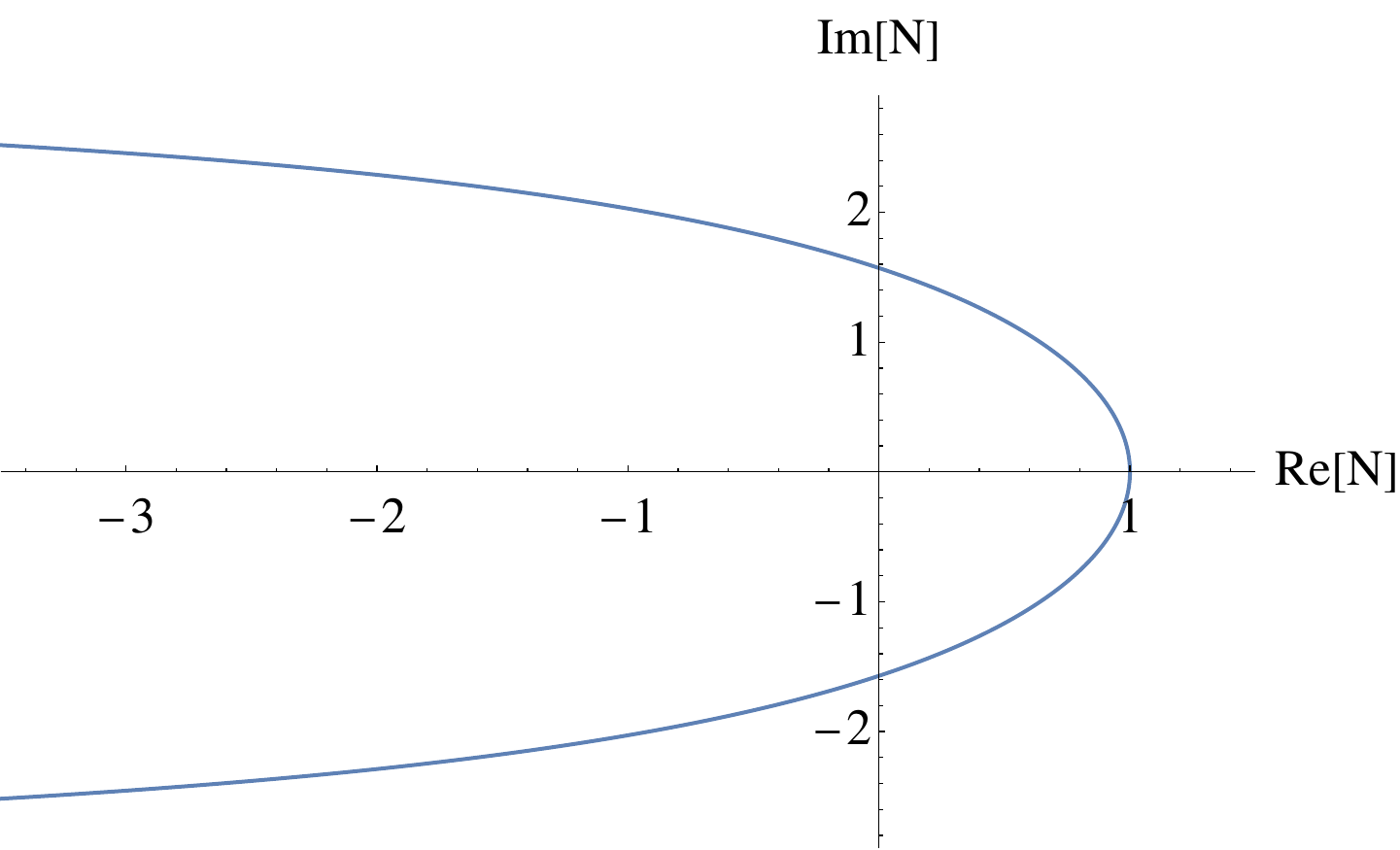}
	\caption{The contour chosen in the Talbot algorithm..
	}
	\label{fig:contour}
\end{figure}
We can choose a contour as
\beq
N=N(\theta)=r\theta(\cot\theta+i),
\eeq 
which is shown in the complex-$N$ plane as Fig. \ref{fig:contour}. The inverse Mellin integration becomes
\begin{equation}
f(x)=\frac{1}{2\pi i}
\int_{-\pi}^{\pi}\dd\theta\frac{\dd N}{\dd\theta}
x^{-N(\theta)}\tilde{f}(N(\theta)).
\end{equation}
With the analytic condition, $\tilde{f}(N^*)=\tilde{f}^{*}(N)$, we can simplify it as 
\begin{equation}
f(x)=\frac{r}{\pi}\int_0^\pi\dd\theta 
\Re{x^{-N(\theta)}\tilde{f}(N(\theta))[1+i\sigma(\theta)]},
\end{equation}
where
\begin{equation}
\dd N/\dd\theta=ir[1+i\sigma(\theta)], ~\sigma(\theta)=\theta+(\theta\cot\theta-1)\cot\theta.
\end{equation}
This integration can be computed with the a trapezoidal rule:
\begin{equation}\label{eq:Talbot}
f(x)\simeq \frac{r}{K}\left[
\frac{1}{2}x^{-r}\tilde{f}(r)+\sum_{k=1}^{K}\Re{
	x^{-N(\theta_k)}\tilde{f}(N(\theta_k))[1+i\sigma(\theta_k)]}
\right],
\end{equation}
where $\theta_k=k\pi/K$.
With an optional choice suggested by Ref. \cite{Abate:2004a},
\begin{equation}
r=\frac{2K}{5\log(1/x)},
\end{equation}
the relative precision can approximately reach $10^{-0.6K}$.

We have validated the Talbot algorithm with the approach of straight line contours developed in Ref. \cite{Vogt:2004ns}.

\begin{figure}
	\includegraphics[width=0.45\textwidth]{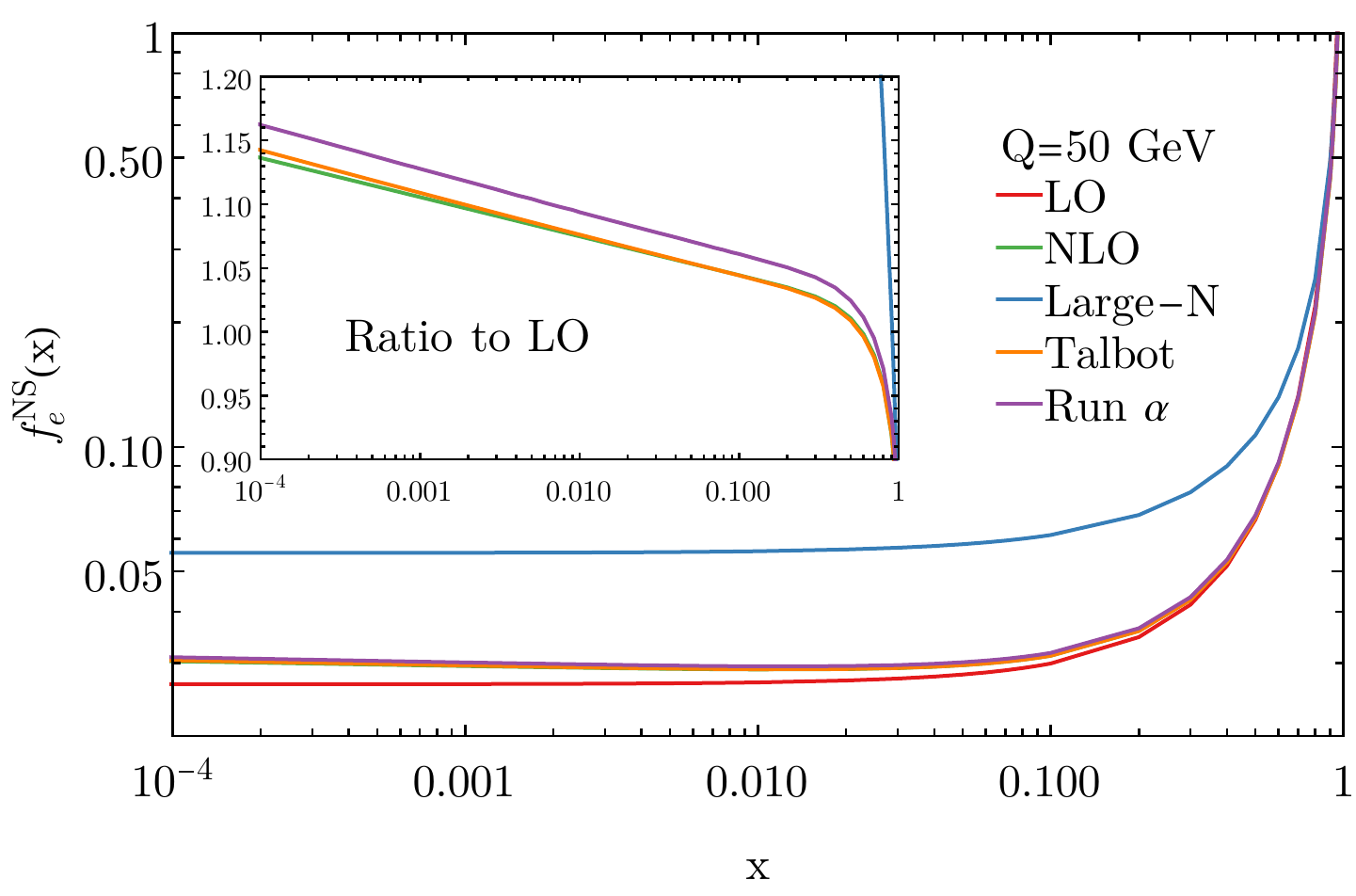}
	\includegraphics[width=0.45\textwidth]{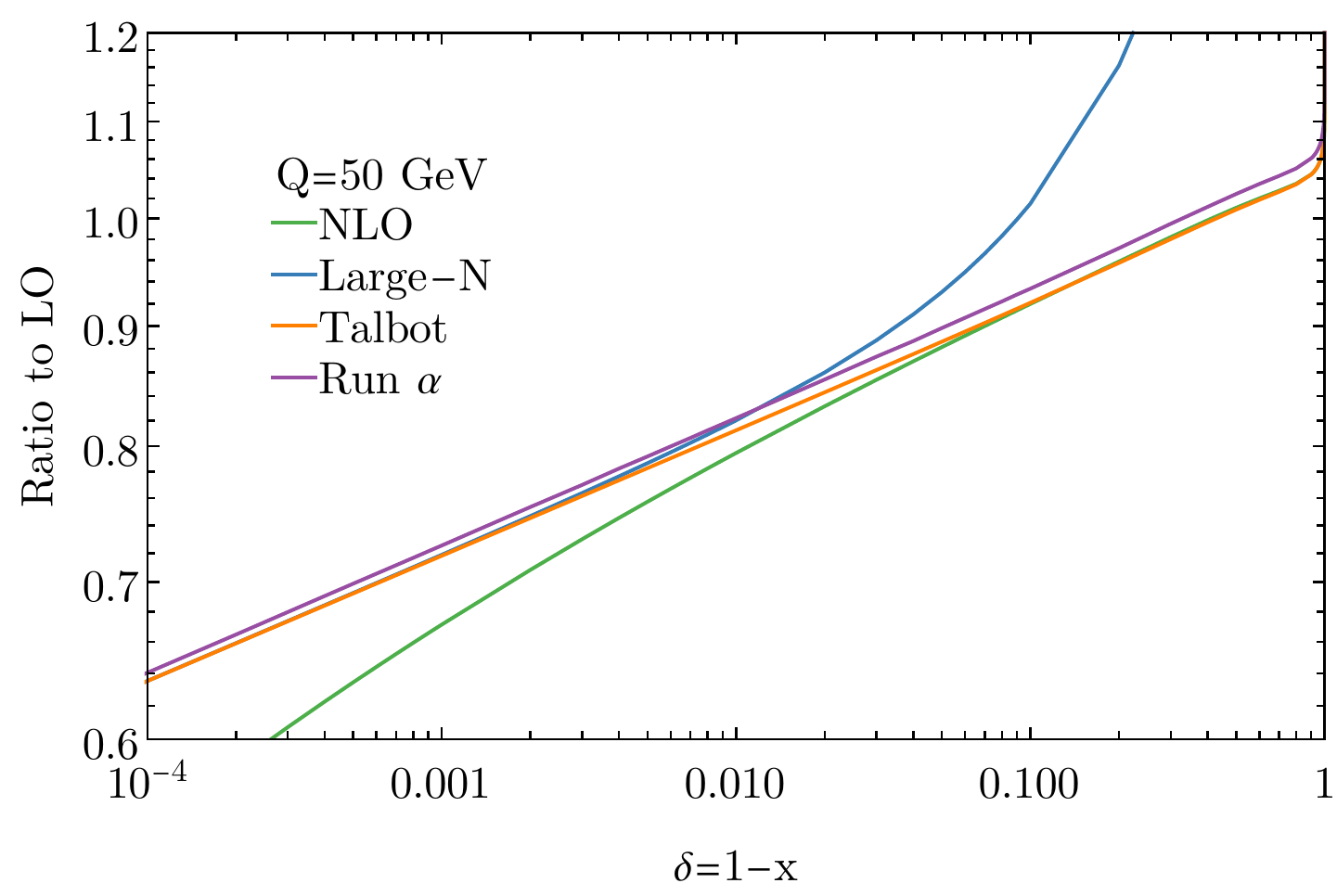}	
	\caption{The non-singlet PDF $f^{\rm NS}_{e}(x)$ at $Q=50$ GeV. The LO and NLO denote perturbative solutions to DGLAP equation at the leading and next-to-leading orders. The ``Large-N" refers to the large Mellin-$N$ approximation in Eq. (\ref{eq:large-x}). The ``Talbot" corresponds to the numerical solution with our iteration and Talbot inversion techniques, while ``Run $\alpha$" extends this approach to incorporate the running coupling $\alpha$ as well. The right plot is the same quantity normalized to the LO solution, with respect to $\delta=1-x$.}
	\label{fig:feNS}
\end{figure}

{\bf Demonstration.}
With the numerical techniques we develop above, we explicitly demonstrate the solution to non-singlet PDF $f_{e}^{\rm NS}(x)$ at $Q=50~\GeV$ in Fig. \ref{fig:feNS}. As a first step, we fix the QED coupling as $\alpha=1/137$. 
With the perturbative expansion, we can obtain the leading and next-to-leading order solutions as
\bea
f_e^{\rm NS,LO}(x,Q^2)&=\delta(1-x)+LP_{\ell\ell},\\
f_e^{\rm NS,NLO}(x,Q^2)&=f_e^{\rm NS,LO}(x,Q)+\frac{1}{2}L^2P_{\ell\ell}\otimes P_{\ell\ell},
\eea
which are shown as the red and green lines in Fig. \ref{fig:feNS}. With normalized to the LO result shown as the inset, we see the NLO receives positive (negative) correction in the small (large) $x$ region. We show the large-$N$ approximation of Eq. (\ref{eq:large-x}) as the blue line and the one with our iteration and Talbot inversion (denoted as ``Talbot") as the orange line. On the right plot, we show the ratio of the LO solution in terms of $\delta=1-x$ to highlight the large $x$ region. We see the Talbot result approaches to the large-$N$ solution when $x\to1$. At a moderate $x$, such as $x<0.9$, we see the obvious deviation of the large-$N$ solution from the Talbot one, indicating the limitation of the large-$N$ approximation. Instead, in the small $x$ limit, we see the Talbot coincides with the NLO solution, with a very small correction, shown as the difference between green and orange lines. It implies the fast convergence of the expansion approach, as mentioned in the last subsection. 

Finally, we also extend the iteration and Talbot inversion approach with a running QED coupling in Eq. (\ref{eq:runAlfa}) and show the results as the purple lines in Fig. \ref{fig:feNS}. Due to the increase of $\alpha$ with scale, we obtain an enhancement at a few percent level for $f_{e}^{\rm NS}(x)$ at $Q=50~\GeV$. The ability to deal with the running coupling is very critical and will make a significant difference when we solve the matrix equation of singlet, photon, and gluon PDFs, which involves the running QCD coupling. The details are left for a future work \cite{ours}.

\bibliographystyle{JHEP}
\bibliography{ref}

\end{document}